\documentclass[twocolumn]{aastex631}

\usepackage{amsmath,amssymb}
\usepackage{bm} 
\usepackage{bbold} 
\usepackage{todonotes}
\usepackage{soul} 

\newcommand{\vhat}[1]{\ensuremath{\widehat{\bm{#1}}}}%
\newcommand{\uvct}[1]{\hat{\mathrm{\mathbf{#1}}}}
\newcommand{\vct}[1]{\bm #1}
\newcommand{\Up}[3]{U^{#1}_{#2, #3}}
\newcommand{\avg}[1]{\left\langle #1 \right\rangle}
\newcommand{\dd}[0]{\ensuremath{\mathrm{d}}}%
\newcommand{\ee}[0]{\ensuremath{\mathrm{e}}}%
\newcommand*{\ii}{\imath}
\newcommand{\wignerj}[6]{\begin{pmatrix} #1 & #2 & #3 \\ #4 & #5 & #6\end{pmatrix}}

\shorttitle{Transition between ballistic and diffusive}
\shortauthors{Kuhlen et al.}
\begin{document}

\hfill TTK-21-43

\title{No longer ballistic, not yet diffusive--the formation of cosmic ray small-scale anisotropies}

\author[0000-0002-0454-6823]{Marco Kuhlen}
\affiliation{Institute for Theoretical Particle Physics and Cosmology \\
Sommerfeldstr.\ 16, RWTH Aachen University \\
52074 Aachen, Germany}

\author[0000-0002-5611-095X]{Vo Hong Minh Phan}
\affiliation{Institute for Theoretical Particle Physics and Cosmology \\
Sommerfeldstr.\ 16, RWTH Aachen University \\
52074 Aachen, Germany}

\author[0000-0002-2197-3421]{Philipp Mertsch}
\affiliation{Institute for Theoretical Particle Physics and Cosmology \\
Sommerfeldstr.\ 16, RWTH Aachen University \\
52074 Aachen, Germany}

\begin{abstract}
The arrival directions of TeV-PeV cosmic rays are remarkably uniform due to the isotropization of their directions by scattering on turbulent magnetic fields. Small anisotropies can exist in standard diffusion models, however, only on the largest angular scales. Yet, high-statistics observatories like IceCube and HAWC have found significant deviations from isotropy down to small angular scales. Here, we explain the formation of small-scale anisotropies by considering pairs of cosmic rays that get correlated by their transport through the same realisation of the turbulent magnetic field. We argue that the formation of small-scale anisotropies is the reflection of the particular realisation of the turbulent magnetic field experienced by cosmic rays on time scales intermediate between the early, ballistic regime and the late, diffusive regime. We approach this problem in two different ways: First, we run test particle simulations in synthetic turbulence, covering for the first time the TV rigidities of observations with realistic turbulence parameters. Second, we extend the recently introduced mixing matrix approach and determine the steady-state angular power spectrum. Throughout, we adopt magneto-static, slab-like turbulence. We find excellent agreement between the predicted angular power spectra in both approaches over a large range of rigidities. In the future, measurements of small-scale anisotropies will be valuable in constraining the nature of the turbulent magnetic field in our Galactic neighborhood.
\end{abstract}


\section{Introduction}
\label{sec:intro}

The arrival directions of Galactic cosmic rays (CRs) are highly isotropic due to their interactions with interstellar turbulent magnetic fields. This random scattering process effectively isotropizes the arrival directions of CRs and leads to their diffusive transport~\citep{1964ocr..book.....G}. However, current high-statistics observatories like IceCube~\citep{2019ApJ...871...96A} and HAWC~\citep{2018ApJ...865...57A} have observed significant deviations from isotropy down to angular scales of $10^\circ$ at energies between a few TeV and a few PeV. These deviations from isotropy can be quantified by the angular power spectrum, which is defined as 
\begin{equation}
  C_\ell(t) = \frac{1}{4\pi}\int\mathrm{d}\uvct{p}_A\int\mathrm{d}\uvct{p}_B P_\ell(\uvct{p}_A\cdot\uvct{p}_B)f_A(t)f_B(t) \, . \label{eqn:def_APS}
\end{equation}
Here, $\uvct{p}_i=\vct{p}_i/|\vct{p}_i|$ are unit momentum vectors, $P_\ell$ are the Legendre polynomials of degree $\ell$, and \mbox{$f_i(t) \equiv f(\vct{r}_\oplus,\vct{p}_i,t)$} is the phase-space density measured by an observer at position $\vct{r}_\oplus$ and time $t$ with momentum $\vct{p}_i$.

In fact, small-scale anisotropies have been the subject of intensive research not only observationally but also theoretically. Many different scenarios have been proposed to explain the presence of anisotropies at small angular scales which includes, for example, effects of the heliosphere~\citep{Drury:2013uka,2014ApJ...790....5Z,2014Sci...343..988S}, non-uniform pitch-angle diffusion~\citep{2010ApJ...721..750M}, non-diffusive propagation of Galactic CRs~\citep{battaner2011,harding2016}, and also more exotic explanations~\citep{kotera2013}.

Interestingly, it has been suggested that the turbulent Galactic magnetic field could be the source of small-scale anisotropies \citep{giacinti2012,2014PhRvL.112b1101A,2015ApJ...815L...2A,battaner2015,lopez-barquero2016,2019JCAP...11..048M}. This idea has been investigated using test particle simulations of CRs in synthetic turbulence where sky maps of the CR arrival directions can be obtained from the phase-space density back-tracked along the CR trajectories to an earlier time (see \citealt{2017PrPNP..94..184A} for a review). Most of the previous studies, however, limit themselves to particles with a ratio of gyroradius to correlation length $\rho \equiv r_g/L_c \sim 10^{-2}-10^{1}$. This corresponds to energies much larger than what is relevant for observational data, e.g.\ $\rho \sim 10^{-4}$, for a $10\,\text{TeV}$ particle in a $4\,\mu \text{G}$ RMS magnetic field of correlation length $L_c \simeq 30\,\text{pc}$. Comparison with observed anisotropies is thus based on some extrapolations in energy. 

More importantly, the standard picture of CR transport known as Quasi-Linear Theory (QLT)~\citep{1966ApJ...146..480J,1966PhFl....9.2377K,1967PhFl...10.2620H,1970ApJ...162.1049H} fails to explain the observed anisotropies at scales smaller than the dipole. This can be traced back to the fact that QLT allows computing only the phase-space density averaged over an ensemble of turbulent magnetic fields, $\langle f\rangle$. Computing the ensemble-averaged angular power spectrum via Eq.~\eqref{eqn:def_APS}, however, requires the two-point function of phase-space densities, $\langle f_A f_B\rangle$. Use of QLT therefore relies on assuming statistical independence, $\langle f_Af_B\rangle = \langle f_A\rangle\langle f_B\rangle$.

Correlations are expected to be present, however: Imagine two particles arriving under a small angle. For about a scattering time before observation, they have experienced the same magnetic field configuration. Fluxes from nearby directions will therefore be similar and $\langle f_A f_B \rangle \geq \langle f_A\rangle \langle f_B \rangle$. \citet{2019JCAP...11..048M} take into account these correlations and put forward a model to predict the angular power spectrum based on a perturbative expansion of the time-evolution operator. However, a rather unrealistic white-noise power spectrum of turbulence was adopted to allow for some explicit analytical results.

The aim of this work is, thus, to further improve our understanding of turbulence-induced small-scale anisotropies. A better understanding of these small-scale anisotropies will allow deriving independent constraints on the nature of turbulence in our Galactic neighborhood, e.g.\ on the CR scattering time or the power spectrum of the local turbulent magnetic field. These quantities are crucial in achieving a better precision of predictions of CR fluxes, but currently, they are only poorly constrained. 

The outline of this paper is as follows. In Sec.~\ref{sec:method}, we will outline the methodological approach which is both numerical and analytical. Sec.~\ref{sec:numerical} introduces the concept of backtracking and gives some details on the numerical test particle simulations. Sec.~\ref{sec:analytical} contains the analytical framework introduced by \citet{2019JCAP...11..048M} for computing the angular power spectrum, extended to the case of slab turbulence. In Sec.~\ref{sec:results}, we will present the numerical and analytical results and compare the computed angular power spectra in the energy range relevant to observations. We will summarise our findings and conclude in Sec.~\ref{sec:summary}.

\section{Method}
\label{sec:method}

\subsection{Numerical simulations}
\label{sec:numerical}

\subsubsection{The concept of backtracking}

The origin of small-scale anisotropies can be understood by exploiting Liouville's theorem that states that the phase-space density is conserved along particle trajectories through phase-space~\citep{Goldstein:2001}. We can therefore evaluate the phase-space density for an arbitrary point $\{\vct{r}(t), \vct{p}(t)\}$ at time $t'=t$, $f(\vct{r}(t), \vct{p}(t), t)$, if we know the phase-space density at another time $t' = t_0$ along the trajectory $\{\vct{r}(t'), \vct{p}(t')\}$ from that initial point,
\begin{equation}
f(\vct{r}(t), \vct{p}(t), t) = f(\vct{r}(t_0), \vct{p}(t_0), t_0) \, .
\label{eq:Liouville}
\end{equation}
Applied to CRs, out of all possible trajectories we pick those that converge at the observer's position $\vct{r}(t) = \vct{r}_{\oplus}$ with $\vct{p}(t) = p \uvct{p}_{\oplus}$ at time $t$. (For simplicity, we have assumed that the absolute value of the momentum $p = |\vct{p}|$ does not change which is a very good approximation for CR nuclei of rigidity above a few GV.) The angular dependence in $f(\vct{r}_{\oplus}, p \uvct{p}_{\oplus}, t)$ then constitutes the CR sky map. This so-called Liouville mapping or back-tracking is also the basis for the computation of the angular power spectrum from test particle simulations of CRs in synthetic turbulence, to be presented in Sec.~\ref{sec:simulations}. We assume the phase-space density to be the quasi-steady solution of a diffusion equation at the earlier time $t_0$ (see e.g. \citealt{2017PrPNP..94..184A}),
\begin{equation}
\avg{ f(\vct{r}(t_0),\uvct{p}(t_0),t_0) } = \bar{f} + \mathbf{r}(t_0)\cdot \nabla\bar{f} - 3\hat{\mathbf{p}}(t_0)\cdot \mathbf{K}\cdot \nabla \bar{f} \, ,
\label{eqn:quasi_steady}
\end{equation}
parametrized by a time-independent isotropic part $\bar{f}$ and its gradient, $\nabla\bar{f}$. Note that in general the phase-space density $f$ in a particular realisation of the turbulent magnetic field differs from the ensemble average, $\avg{ f }$, that is $\delta f \equiv f - \avg{ f } \neq 0$. In backtracking, however, this is negligible since $\delta f(t_0)$ is uncorrelated with $f(t)$ as long as the backtracking time $T \equiv (t - t_0)$ is large enough.

\begin{figure*}[tbh]
\includegraphics[scale=1]{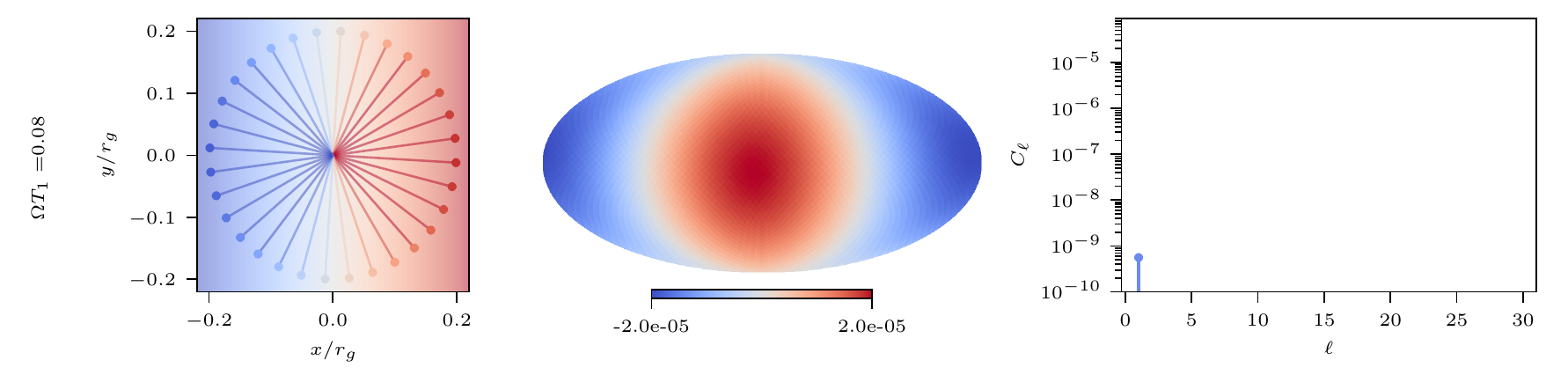}\\
\includegraphics[scale=1]{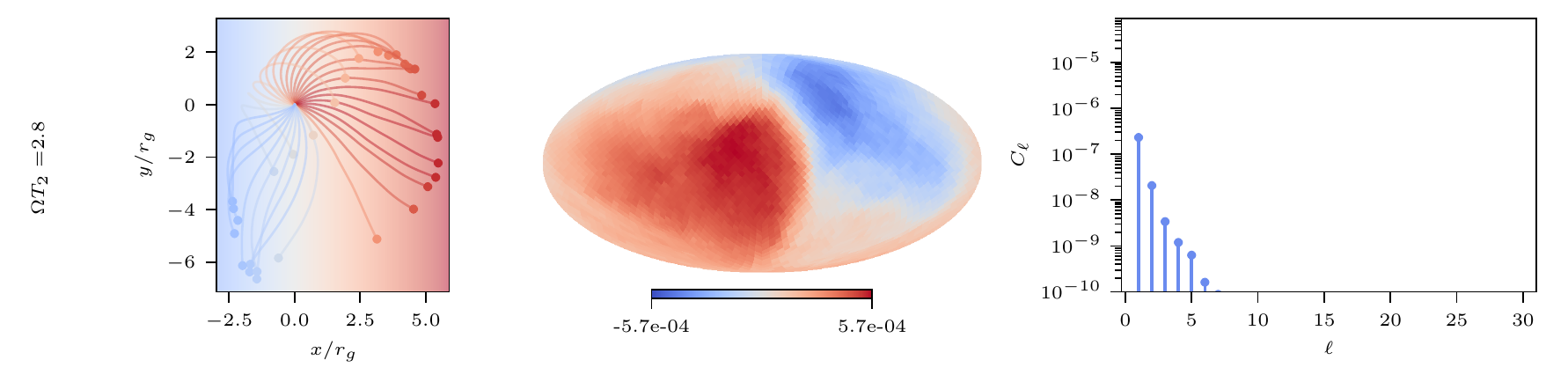}\\
\includegraphics[scale=1]{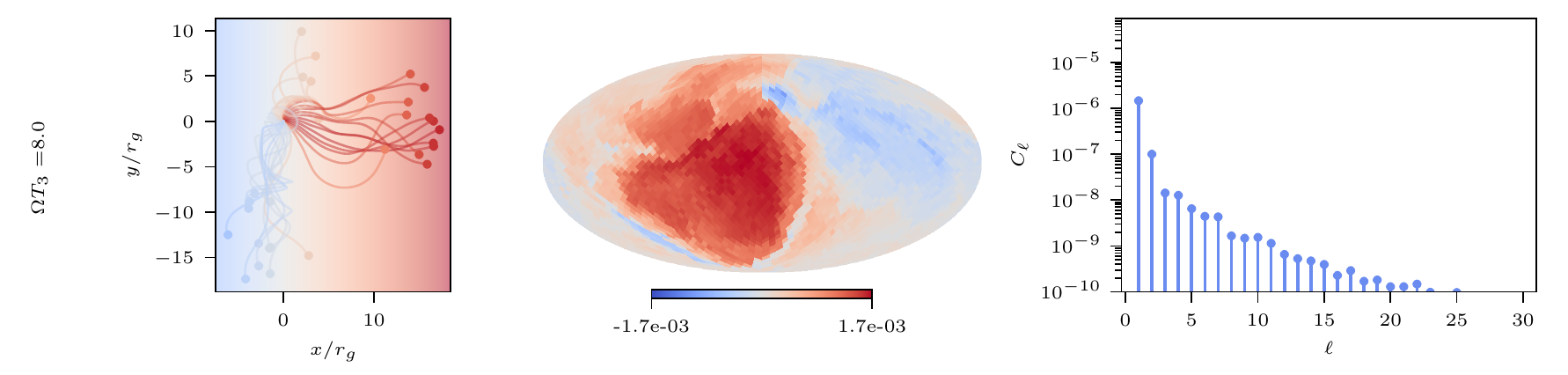}\\
\includegraphics[scale=1]{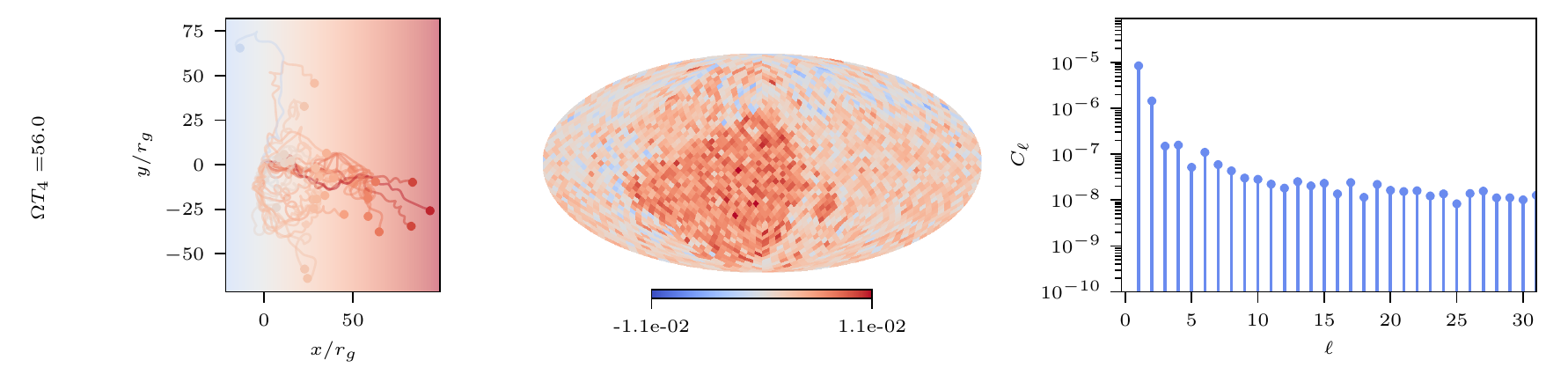}
\caption{Trajectories (left column), sky maps (middle column) and angular power spectra (right column) for different times $T$ (top through bottom row). The times $T_1$ through $T_4$ can be considered as backtracking times $(t - t_0)$ when following particles back in time along their trajectories or as the times passed since the preparation of the initial phase-space density at $t_0$. See the text for further explanations.}
\label{fig:illustration}
\end{figure*}

For illustration, we follow a set of trajectories, their sky map and its angular power spectrum as we increase the backtracking time $T$ from $T_1$ to $T_2$, $T_3$ and $T_4$ as shown in Fig.~\ref{fig:illustration}. Backtracking for a small time $T_1$ only (top row), CRs have travelled mostly ballistically, their trajectories being straight lines (top left panel). The phase-space density at the other end of the trajectory $i$ at time $t_0 = t-T_1$ and therefore the flux in a direction $\vhat{p}_i$ at time $t$ is proportional to the projection of this direction onto the gradient $\vct{\nabla} \bar{f}$, $f(\vct{r}_{\oplus}, \vct{p}_i, t) \propto \uvct{p}_i \cdot \vct{\nabla} \bar{f}$. The sky map (top center panel) is then approximately dipolar and the angular power spectrum $C_{\ell}$ (top right panel) vanishes for $\ell > 1$.

If we backtrack for a longer time $T_2 > T_1$ (second row from top), we can see (left panel) that trajectories that arrive at the observer from nearby directions at time $t$ were mostly already closely aligned a time $T_2$ earlier: For instance, a number of trajectories towards the top end of the panel have just started gyrating around the large-scale magnetic field. Another set of trajectories, towards the right side of the panel are instead moving towards the right without much gyration since their momentum vectors are aligned with the background magnetic field. Therefore, the phase-space densities at the ends of the trajectories are correlated for trajectories from nearby observed directions. The sky map (centre panel), while not perfect dipolar anymore, is still rather smooth. There is some angular power (right panel) on scales different from the dipole, but the smaller scales are still largely suppressed.

For even larger backtracking times $T_3 > T_2$ (second row from top), the trajectories (left panel) have further differentiated out, but nearby trajectories are still correlated. On the sky map (centre panel), this is visible as structure on smaller scales. Correspondingly, we now observe power on almost all scales in the angular power spectrum (right panel).

Finally, for very late times $T_4 > T_3$ (bottom row), the particle directions and positions are finally uncorrelated. In the sky map (centre panel) this is reflected in noise on small angular scales, also visible as a flat angular power spectrum for $\ell \gtrsim 10$. We are thus led to conclude that the specific, non-dipolar pattern is imprinted onto the map of CR arrival directions at intermediate times, between the epochs of purely ballistic and of purely diffusive transport. This transition has been the topic of recent interest in the study of CR transport~\citep[e.g.][]{2014ApJ...783...15E,2015PhRvD..92h3003P,2017PhRvD..95b3007M}.

So far, we have read the succession of the trajectories, sky maps and angular power spectra in Fig.~\ref{fig:illustration} from top to bottom as going successively further in backtracking time. It is however perfectly equivalent to read them from top to bottom as going forward in the time that has passed since preparing the phase-space density of Eq.~\eqref{eqn:quasi_steady}. Imagine we \emph{globally} prepare the phase-space density at time $t_0$ to Eq.~\eqref{eqn:quasi_steady} and then observe its evolution at the observer's position. The particles arriving at time $(t_0 + T_1)$ will have been at the positions indicated by the markers in the top left panel of Fig.~\ref{fig:illustration} at time $t_0$. Given Liouville's theorem, the phase-space densities at these points have been conserved along the individual trajectories and are observed at time $(t_0 + T_1)$ as the sky map (middle panel). Correspondingly, the right panel shows the angular power spectrum a time $T_1$ after setting up the initial state. Similarly, the particles arriving at time $(t_0 + T_2)$ (second row) were at the end of the trajectories shown in the left panel at time $t_0$. The sky map observed at time $(t_0 + T_2)$ and its angular power spectrum are shown in the middle and right panels. The same applies to the other rows for times $(t_0 + T_3)$ and $(t_0 + T_4)$.

The succession of sky maps and angular power spectra from top (time $t_0 + T_1$) to bottom (time $t_0 + T_4$) therefore shows the evolution of angular structures as seen by the observer as time progresses. Initially, all the angular power is on large scales, but at later times power starts appearing on smaller scales. While the processes of this generation has a \emph{non-local} character (due to the Liouville mapping), in the ensemble average we can describe it \emph{locally} as an evolution of angular power from large scales to smaller scales. Given the discussion above, it is clear that the angular power on small scales is produced by mixing of power from larger into smaller scales: Structure coherent on a certain angular scale starts developing substructure on smaller scales as the set of trajectories corresponding to the initial scale start diverging.

\subsubsection{Synthetic turbulence}
\label{sec:simulations}

We start by numerically simulating test particles in synthetic turbulence~\citep[e.g.][]{2020Ap&SS.365..135M}. The monoenergetic test particles are initialized 
with isotropic directions on a HEALPix~\citep{2005ApJ...622..759G} grid with $N_\text{side} = 256~\mathrm{or}~512$ leading to a total number of test particles $N_\text{particles} \sim 8\times 10^5~\text{and}~3\times 10^6$. These particles are then tracked back in time through the magnetic field by solving the Newton-Lorentz equations using the rigidity conserving Boris method~\citep{boris}. As the test particles do not interact with each other or backreact on the magnetic field this can be parallelised very efficiently. We therefore run these simulations on GPUs which allow for very cost-effective parallelisation \citep{2016NewA...45....1T,GPU-lecturenotes}.

In the literature two different methods have been used to generate synthetic magnetic field turbulence. In the method proposed by \citet{1994ApJ...430L.137G}, the magnetic field is calculated as a superposition of waves. Only the phases and amplitudes for the waves are stored. In the other method the magnetic field is set up on a grid in Fourier space, transformed to and saved in real space. This has the advantage that no large sums have to be evaluated at every particle position and time step. The magnetic field evaluation is reduced to a simple interpolation between grid points \citep{Schlegel:2019uww}. The disadvantage is the large amount of memory required to store the entire field grid.

The rigidity of particles required to compare to observational data from IceCube and HAWC is of the order of $10\,\text{TV}$ assuming only protons. Therefore the resulting gyroradii that need to be resolved in our simulations are \mbox{$r_g/L_c\sim 8.3\times 10^{-5}$} given $B_{\rm RMS}\simeq 4\,\mu$G  and $L_c\simeq 30 \, \text{pc}$. Even though there could be artefacts due to grid periodicity, it has proven sufficient to make the grid a factor of  $8$ larger than the largest wavelength. To get the correct normalization of the angular power spectrum the smallest wavelength has to be much smaller than the gyroradius. In practice we choose the smallest wavelength to be a factor $\sim 150$ smaller than the gyroradius at the smallest rigidity and resolve it by at least 10 grid points. Spanning this large dynamical range with a single grid would require at least $n=675,000,000$ grid points. To reduce the memory requirement on our GPUs we therefore use 3 nested grids    with different grid spacings as proposed by~\citet{2012JCAP...07..031G}. Each individual grid covers a part of the magnetic field power spectrum. The 3 grids are then superimposed \citep[see][for an illustration]{2020Ap&SS.365..135M}. 

Equation \eqref{eq:Liouville} is applied for the simulated CR trajectories to provide a sky map for a particular realization of turbulence. We then calculate and average the angular power spectra for all realizations to obtain the ensemble-averaged numerical angular power spectrum $\langle C_\ell\rangle^{\rm num}$. At this point, we redefine the angular power spectrum from Eq.~\eqref{eqn:def_APS} as follows 
\begin{equation}
  \langle C_\ell\rangle = \frac{1}{4\pi}\int\mathrm{d}\uvct{p}_A\int\mathrm{d}\uvct{p}_B P_\ell(\uvct{p}_A\cdot\uvct{p}_B)\frac{\langle f_Af_B\rangle}{\bar{f}^2} \, . \label{eqn:redef_APS}
\end{equation}
which essentially replaces the phase-space density by the relative intensity of CRs as commonly adopted for the observed angular power spectrum. 

We note also that even for a large number of particles the higher multipoles of the angular power spectrum are strongly effected by noise. For large backtracking times the particle transport is dominated by diffusion. The sky map therefore gets dominated by Gaussian noise with variance 
\begin{equation}
\sigma^2 =\langle r_i r_j\rangle \frac{\partial_i \bar{f}\partial_j\bar{f}}{\bar{f}^2}= 2T K_{ij}\frac{\partial_i \bar{f}\partial_j\bar{f}}{\bar{f}^2}.
\label{eqn:sky_variance}
\end{equation}

The angular power spectrum coming from this noise contribution can be calculated by substituting \mbox{$\langle f_A f_B \rangle = \sigma^2 \delta^{(3)} (\uvct{p}_A - \uvct{p}_B)$} into the definition of the angular power spectrum given in Eq.~\eqref{eqn:redef_APS}. In the case of a discrete pixelized sky map this gives
\begin{equation}
    \mathcal{N}_\ell = \frac{1}{4\pi}\sum_i \Delta\Omega \sum_j \Delta \Omega P_\ell(\uvct{p}_i\cdot\uvct{p}_j) \sigma^2 \delta_{ij},
\end{equation}
with the pixel size $\Delta \Omega = 4\pi/N_\text{pix}$, where $N_\text{pix}$ is the total number of pixels in the arrival direction sky map. 
Evaluating the angular power spectrum of the noise contribution, using that $P_\ell(1) = 1$, results in a constant contribution for all $\ell$ that is given by~\citep{2015ApJ...815L...2A}

\begin{equation}
\mathcal{N}_\ell = \frac{4\pi}{N_\text{pix}} 2 T K_{ij} \frac{\partial_i \bar{f}\partial_j \bar{f}}{\bar{f}^2} .\label{eqn:noise}
\end{equation}
It has been argued by \citet{2015ApJ...815L...2A} that the best estimator for the angular power spectrum in this case is $\langle C_\ell\rangle^{\rm sub}=\langle C_\ell\rangle^{\rm num} -\mathcal{N}_\ell$ which we refer to throughout this work as the ensemble-averaged, noise-subtracted angular power spectrum. The variance of $\langle C_\ell\rangle^{\rm sub}$ can be estimated as $2\mathcal{N}_\ell^2/(2\ell+1)$~\citep{2015MNRAS.448.2854C}.

\subsubsection{Parameter values}
\label{sec:parameters}

For all simlations we choose an outer scale of turbulence $L_\text{max} = 150\,\text{pc}$, that was used also in previous studies~\citep[e.g.][]{2018JCAP...07..051G}. For a kolmogorov turbulence like power spectrum this corresponds to a correlation length $L_c \simeq L_\text{max}/5 = 30\,\text{pc}$~\citep{2002JHEP...03..045H}. Current observational constraints on the outer scale differ significantly~\citep{2013A&A...558A..72I} depending on galactic latitude~\citep{2010ApJ...710..853C} and the position in the galactic plane~\citep{2008ApJ...680..362H}. Since the simulations are only sensitive to the normalized rigidity $\rho=r_g/L_c$, simulations at one $\rho$ can correspond to a range of outer scales and rigidities. The total root mean square magnetic field strength is set to $\sqrt{B_0^2+\langle\delta B^2\rangle} = 4\,\mu\text{G}$~\citep{2013pss5.book..641B}, motivated by energy equipartition calculations. The turbulence level indicated by modelling efforts of the galactic magnetic field, based on the observed synchrotron intensity and polarization, $\eta = \langle\delta B^2\rangle/(B_0^2+\langle\delta B^2\rangle) \sim 0.7-0.9$~\citep{2013pss5.book..641B,2017A&A...603A..15O} seems to be incompatible with the CR diffusion coefficient obtained from measurements of the Boron to Carbon ratio~\citep{Evoli:2019wwu,Weinrich:2020cmw} at GeV energies under standard assumptions~\citep{2018JCAP...07..051G}, pointing to much smaller turbulence levels. We will therefore focus on an intermediate turbulence level of $\eta=0.5$. The results of the numerical simulations will be presented in Sec.~\ref{sec:results}.

\subsection{Analytical computation}
\label{sec:analytical}

\subsubsection{The mixing equation}
\label{sec:mixing-equation}

We will start by reviewing the diagrammatic formalism for solving the Liouville equation that describes the time evolution of the phase-space density, as outlined by \citet{2019JCAP...11..048M}. We shall work in natural units ($c=1$) and consider only relativistic particles for simplicity. The Liouville equation for relativistic charged particles propagating through the magnetic field with both static regular and turbulent components, denoted as $\vct{B}_0$ and $\delta\vct{B}$, could be written as 
\begin{equation}
\label{eq:vlasov}
  \partial_t f + \uvct{p}\cdot\vct{\nabla} f + \mathcal{L} f = -\delta \mathcal{L} f \, ,
\end{equation}
where $\mathcal{L} = -\ii \vct{\Omega} \cdot \vct{L}$ and $\delta\mathcal{L} = - \ii \vct{\omega} \cdot \vct{L}$ represent the deterministic and stochastic Liouville operators. Here, $\vct{\Omega} = q \vct{B}_0/p$ and $\vct{\omega} = q\delta\vct{B}/p$ are the gyrovectors in the regular and turbulent magnetic fields, respectively. The $L_m=\ii\varepsilon_{mnk}p_n\partial/\partial p_{k}$ are angular momentum operators that obey the usual commutation relations $[ L_m, L_n ] = \ii \epsilon_{mnk} L_k$. 

In the absence of any spatial dependence, $\vct\nabla f \equiv 0$, this equation can be solved formally,
\begin{equation}
f(\vct{p}, t) = U_{t,t_0} f(\vct{p}, t_0) \, ,
\end{equation}
by introducing a time evolution operator or propagator
\begin{equation}
U_{t,t_0} = \mathcal{T}\exp{\left[-\int_{t_0}^{t}\mathrm{d}t'(\mathcal{L}+\delta\mathcal{L}(t'))\right]} \, , \label{eqn:time_evolution_operator}
\end{equation}
with the usual time-ordered ("latest-to-left") exponential. The time evolution operator depends, through $\delta\mathcal{L}$, on the specific realisation of the turbulent magnetic field, $\vct{\omega}$ to which we do not have access in any practical sense. What we can predict, however, are statistical averages. The ensemble averaged phases space density $\langle f \rangle$, for instance satisfies,
\begin{equation}
\langle f(\vct{p}, t) \rangle = \langle U_{t,t_0} f(\vct{p}, t_0) \rangle \simeq \langle U_{t,t_0} \rangle \langle f(\vct{p}, t_0) \rangle \, .
\end{equation}
In the last step, we have assumed that $(t - t_0)$ is so large, that correlations between the initial state $f(\vct{p}, t_0)$ and the time evolution operator $U_{t,t_0}$ can be ignored.

Relaxing the requirement of homogeneity, we can expand the phase-space density around the position of the observer $\vct{r}_\oplus$ as $f(\vct{r},\vct{p},t) \simeq f_\oplus + (\vct{r} - \vct{r}_\oplus) \cdot \bm\nabla\bar{f}$, where $\bm\nabla\bar{f}$ denotes a CR gradient at $\vct{r}_\oplus$, and insert this into Eq.~\eqref{eq:vlasov} to obtain an equation for the time evolution of the phase-space density at the position of the observer
\begin{equation}
    \label{eq:vlasov2}
    \partial_t f_\oplus + \mathcal{L} f_\oplus + \delta\mathcal{L}f_\oplus \simeq - \uvct{p}\cdot \nabla\bar{f}.
\end{equation}
As we will show later, the right hand side sources a dipole anisotropy coming from the assumption of a CR gradient. The formal solution to Eq.~\eqref{eq:vlasov2} can then be written as
\begin{equation}
    \label{eq:solution}
    \begin{aligned}
    f_\oplus(\vct{p},t) 
    &\simeq U_{t,t_0}f_{\oplus}(\vct{p},t_0)-\int^t_{t_0}\mathrm{d} t' U_{t,t'} \uvct{p}(t) \cdot \vct\nabla \bar{f} \\
    &=U_{t,t_0}f_\oplus(\vct{p},t_0) + \Delta\vct{r}(t_0) \cdot \vct\nabla \bar{f}.
    \end{aligned}
\end{equation}
where we have denoted $\Delta\vct{r}(t_0)=\vct{r}(t_0) - \vct{r}_\oplus$. Note also that we have employed $U_{t,t'} \uvct{p}(t)=\uvct{p}(t')$ for the last identity.

As briefly discussed in Sec.~\ref{sec:intro}, the trajectories of a pair of CRs are expected to stay correlated for a finite amount of time prior to observation which results in the presence of small-scale anisotropies. This means that in order to predict the angular power spectrum, we have to examine the ensemble average for the products of phase-space densities $\left\langle f_\oplus(\vct{p}_A,t) f^*_\oplus(\vct{p}_B,t)\right\rangle$ which from Eq.~\eqref{eq:solution} could be written as
\begin{equation}
\label{eq:formal_solution_product}
\begin{aligned}
\avg{f_A(t) f^*_B(t)} & \simeq \avg{\Up{A}{t}{t_0} \Up{B*}{t}{t_0}} \avg{f_A(t_0) f_B^*(t_0)} \\
& +\avg{ \big( \Delta \vct{r}_A(t_0) \cdot \vct\nabla \bar{f} \big) \Up{B*}{t}{t_0} f^*_B(t_0)} \\ 
& + \avg{ \big( \Delta \vct{r}_B(t_0) \cdot \vct\nabla \bar{f}^* \big) \Up{A}{t}{t_0}f_A(t_0) } \\
& \mkern-9mu + \avg{ \big( \Delta \vct{r}_A(t_0) \cdot \vct\nabla \bar{f} \big) \big( \Delta \vct{r}_B(t_0) \cdot \vct\nabla \bar{f}^* \big) }
\end{aligned}
\end{equation}
where we have again adopted the shorthand notation $f_A(t)=f_\oplus(\vct{p}_A,t)$ and $f_B(t)=f_\oplus(\vct{p}_B,t)$. In addition, we have also assumed that the correlations between the propagators and the initial state $f_\oplus(\vct{p},t_0)$ are negligible. Note that we have used the asterisk symbol to mark the complex conjugates of both functions and operators (including the angular momentum operators).

If we could evaluate the pair propagator $\avg{\Up{A}{t}{t_0} \Up{B*}{t}{t_0}}$, the angular power spectrum could be computed at anytime for a given initial state $f(\vct{p},t_0)$ using \mbox{Eqs. \eqref{eqn:redef_APS} and \eqref{eq:formal_solution_product}}. However, it might be difficult to estimate directly the ensemble average of all the terms on the RHS of Eq.~\eqref{eq:formal_solution_product} for an arbitrary period of time $(t-t_0)$ due to the stochastic nature of particles' trajectories in the turbulent magnetic field. Thus, we shall first limit ourselves to the evolution of the angular power spectrum over a short period of time from $t_0$ to $(t_0+\Delta T)$ with small enough $\Delta T$ such that we could approximate $U_{t,t_0}\simeq 1 + \mathcal{O}(\Delta T)$ and also $\lim_{\Delta T\rightarrow 0}\Delta\vct{r}(t_0)\simeq -\mu\Delta T\uvct{z}$. (Note that we have adopted a coordinate system with the $z$-axis directed along $\vct{B}_0$ for definiteness.) Equation~\eqref{eq:formal_solution_product} in this case becomes
\begin{equation}
\begin{aligned}
& \mkern-18mu \avg{f_A(t_0+\Delta T) f^*_B(t_0+\Delta T)} \\
& \simeq \avg{\Up{A}{t_0+\Delta T}{t_0} \Up{B*}{t_0+\Delta T}{t_0}} \avg{f_A(t_0) f_B^*(t_0)} \\
& + (\mu_A\Delta T\partial_z\bar{f})\avg{f^*_B(t_0)} + (\mu_B\Delta T\partial_z\bar{f})\avg{f_A(t_0)} \\
& + \mathcal{O}(\Delta T^2) \, .
\label{eqn:product-small-time}
\end{aligned}
\end{equation}
We can now integrate both sides of Eq.~\eqref{eqn:product-small-time} and adopt the definition of the angular power spectrum in Eq.~\eqref{eqn:redef_APS} to obtain an equation for the time evolution of the angular power spectrum over a small time step $\Delta T$,
\begin{equation}
\begin{aligned}
&\langle C_\ell(t_0+\Delta T)\rangle \simeq \, \! \frac{1}{4\pi} \! \int\mathrm{d}\uvct{p}_A \! \int\mathrm{d}\uvct{p}_B P_\ell(\uvct{p}_A\cdot\uvct{p}_B) \\
& \times \Bigg[ \avg{ \Up{A}{t_0+\Delta T}{t_0} \Up{B*}{t_0+\Delta T}{t_0} } \frac{\avg{f_A(t_0) f_B^*(t_0)}}{\bar{f}^2} \\
 &+ \Delta T\frac{\partial_z\bar{f}}{\bar{f}^2}\Big( \mu_A\avg{f^*_B(t_0)} + \mu_B\avg{f_A(t_0)} \vphantom{\frac{a}{b}} \Big) \Bigg] \, .
\label{eq:product-small-time}
\end{aligned}
\end{equation}

The expression above could be re-written in a more compact form by introducing the so-called mixing matrix as presented in \citet{2019JCAP...11..048M},
\begin{widetext}
\begin{equation}
M_{\ell\ell_0}(\Delta T) = \frac{1}{4\pi}\int\mathrm{d}\uvct{p}_A\int\mathrm{d}\uvct{p}_B P_\ell(\uvct{p}_A\cdot\uvct{p}_B)\langle U^A_{t_0+\Delta T,t_0}U^{B*}_{t_0+\Delta T,t_0}\rangle \frac{2\ell_0+1}{4\pi}P_{\ell_0}(\uvct{p}_A\cdot\uvct{p}_B).
  \label{eq:mixing-matrix}
\end{equation}
\end{widetext}
Note that by construction the mixing matrix depends only on the small time step $\Delta T$ in the case of static turbulence. We shall further assume that the quasi-stationary solution of the ensemble averaged phase-space density applies, meaning $\avg{f(\uvct{p},t_0)}\simeq\bar{f}-3\uvct{p}\cdot\vct{K}\cdot\nabla\bar{f}$, where $\vct{K}$ is the diffusion tensor. This allows us to rewrite the time evolution of the angular power spectrum over a small time step $\Delta T$ as follows
\begin{equation}
\begin{aligned}
 \langle C_{\ell}(t_0+\Delta T) \rangle  &= \! \sum_{\ell_0 = 0}^{\infty}M_{\ell \ell_0}(\Delta T)C_{\ell_0}(t_0) \\
&+  \delta_{\ell 1}\frac{8\pi}{3}K_{zz}\Delta T\left(\frac{\partial_z\bar{f}}{\bar{f}}\right)^2
\end{aligned}
\label{eq:APS-small-time}
\end{equation}
It is clear from Eq. \eqref{eq:APS-small-time} that the angular power spectrum at some later time $(t_0+\Delta T)$ gets contributions both from the mixing induced by the turbulent magnetic field of the angular power spectrum at earlier time $t_0$ (the first term on the RHS) and the dipole source term (the second term on the RHS). In the following we will suppress the summation symbol and use the Einstein sum convention for the matrix product. 

We can generalize this time evolution equation to an arbitrary time $t = t_0 + n\Delta T$ with $n\geq 1$ by applying Eq. \eqref{eq:APS-small-time} consecutively to evolve the angular power spectrum from $C_\ell(t_0)$ to $C_\ell(t)$, yielding 
\begin{equation}
\begin{aligned}
&\langle C_{\ell}(t)\rangle 
= \left( M_{\ell\ell_0}(\Delta T) \right)^n C_{\ell_0}(t_0) \\
& + \sum_{n_0=1}^{n} \left( M_{\ell\ell_0}(\Delta T) \right)^{n_0-1} \delta_{\ell_0 1} \frac{8\pi}{3} K_{zz} \Delta T \left(\frac{\partial_z\bar{f}}{\bar{f}}\right)^2 \, .
\label{eq:APS-full-time}
\end{aligned}
\end{equation}

It might be instructive to gain some insight into the picture of mixing angular power by considering for example an initial state which has only the dipole meaning $C_\ell(t_0)\neq 0$ only for $\ell=1$ similar to the one illustrated in the right panel of the first row in Fig.~\ref{fig:illustration}. 
As we evolve the system in time, the source term (the last term on the RHS of Eq.~\eqref{eq:APS-full-time}) will continuously provide the angular power to the dipole and the mixing matrix starts to transfer angular power from the dipole to smaller scales corresponding to $\ell>1$. 
This would result in the angular power spectrum as in the right panels in the other rows of Fig. \ref{fig:illustration}. This mixing will persist until the system reaches a steady-state and the angular power spectrum becomes time-independent with angular power on both large and small scales. Note that we are making the analogy between the angular power spectra in Eq.~\eqref{eq:APS-full-time} and in Fig.~\ref{fig:illustration} only for illustration and they are not exactly the same. This is because Eq.~\eqref{eq:APS-full-time} actually provides the ensemble-averaged angular power spectrum while Fig. \ref{fig:illustration} only represents a specific realization of the angular power spectrum. 

Another comment on the nature of the dipole source term is in order. The fact that there is only a dipole source term on the RHS of Eq.~\eqref{eq:APS-small-time} can be traced back to the fact that in the derivation of the ensemble-averaged phase-space density $\avg{ f }$ only the monopole and dipole terms in $\uvct{p}$ are retained~\citep{jones1990}. Source terms with $\ell > 1$ would also appear if the presence of higher order terms in $\uvct{p}$ was considered (see e.g.~\citet{giacinti2017}).

We could now seek the time-independent angular power spectrum by considering the steady-state limit of Eq.~\eqref{eq:APS-small-time} or \eqref{eq:APS-full-time}, that is demanding $C_\ell( t ) = C_\ell (t_0)$ which results in the steady-state equation for the angular power spectrum or \textit{the mixing equation},
\begin{equation}
\label{eq:mm}
\frac{\delta_{\ell\ell_0}-M_{\ell\ell_0}(\Delta T)}{\Delta T} \langle C_{\ell_0}\rangle^{\rm mat} = \frac{8\pi}{3}K_{zz}\left(\frac{\partial_z \bar{f}}{\bar{f}}\right)^2 \delta_{\ell 1}.
\end{equation}
where $\langle C_\ell\rangle^{\rm mat}$ denotes the steady-state ensemble-averaged angular power spectrum from the mixing matrix approach. The right hand side of Eq.~\eqref{eq:mm} is again a dipole term sourced by a CR gradient according to Fick's law. 
The left hand side describes how power from the dipole source is mixed into higher multipoles by the mixing matrix.

At this point, it might be appropriate to start a brief discussion on the physical interpretation of $\Delta T$ which is, in fact, the only parameter required to predict the angular power spectrum for a given turbulence model. Since we first specify $\Delta T$ as a small time step in the derivation of Eq. \eqref{eq:APS-small-time}, one might expect that we should take the limit $\Delta T\rightarrow 0$ in the mixing matrix $M_{\ell \ell_0}$. However, we shall see below that the interactions between particles mediated by the turbulent field are essential for the generation of small-scale anisotropies as they allow the mixing of angular power from large to small scales. It might be, therefore, more suitable to choose $\Delta T$ as a timescale on which these interactions occur. In practice, this means that the time integration of the pair propagators in Eq. \eqref{eq:mixing-matrix} for the mixing should be carried out for a finite amount of time. Determining the correct value of $\Delta T$ requires further examination of the time evolution operators and will be done later in Sec.~\ref{sec:OmegaDT}. Nonetheless it is important to note at this point that there are two constraints on this parameter: i) $\Omega\Delta T \gtrsim 2\pi$, to allow for gyroresonant interactions and ii) $\Delta T < \tau_s$ where $\tau_s=K_{zz}/3$ is the scattering time in QLT (setting again $c=1$ in natural unit). The latter constraint is due to the fact that we have assumed unperturbed trajectories in the derivation of the mixing equation.

\subsubsection{The mixing matrix}
\label{sec:mixing_matrix}

In order to compute the mixing matrix $M_{\ell\ell_0}$, we need to expand the time-ordered exponential of Eq.~\eqref{eqn:time_evolution_operator} into an infinite series and compute the expectation values of the individual terms. Note that under the assumption of Gaussianity, the $n$-point functions in the turbulent magnetic field $\vct{\omega}$ only contribute for even $n$. Further, through a cumulant expansion~\citep{1962JPSJ...17.1100K} they can be expressed as sums of permutations of products of $2$-point functions, e.g.,
\begin{equation*}
\begin{aligned}
\langle \vct{\omega}(t_1) \vct{\omega}(t_2) \vct{\omega}(t_3) \vct{\omega}(t_4) \rangle 
&\!=\! \langle \vct{\omega}(t_1) \vct{\omega}(t_2) \rangle \langle \vct{\omega}(t_3) \vct{\omega}(t_4) \rangle \\
&\!+\! \langle \vct{\omega}(t_1) \vct{\omega}(t_3) \rangle \langle \vct{\omega}(t_2) \vct{\omega}(t_4) \rangle \\
&\!+\! \langle \vct{\omega}(t_1) \vct{\omega}(t_4) \rangle \langle \vct{\omega}(t_2) \vct{\omega}(t_3) \rangle \, .
\end{aligned}
\end{equation*}

In fact, the Bethe-Salpeter equation \citep{salpeter1951} allows us to expand the pair propagator $\avg{\Up{A}{t}{t_0} \Up{B*}{t}{t_0}}$ in Eq.~\eqref{eq:formal_solution_product} into a perturbative series in powers of the turbulent magnetic field which could be conveniently described with a diagrammatic representation similar to Feynman diagrams in Quantum Field Theory,
\begin{equation}
\begin{matrix}\includegraphics[scale=1.0,trim={1cm 0cm 1cm 0cm}]{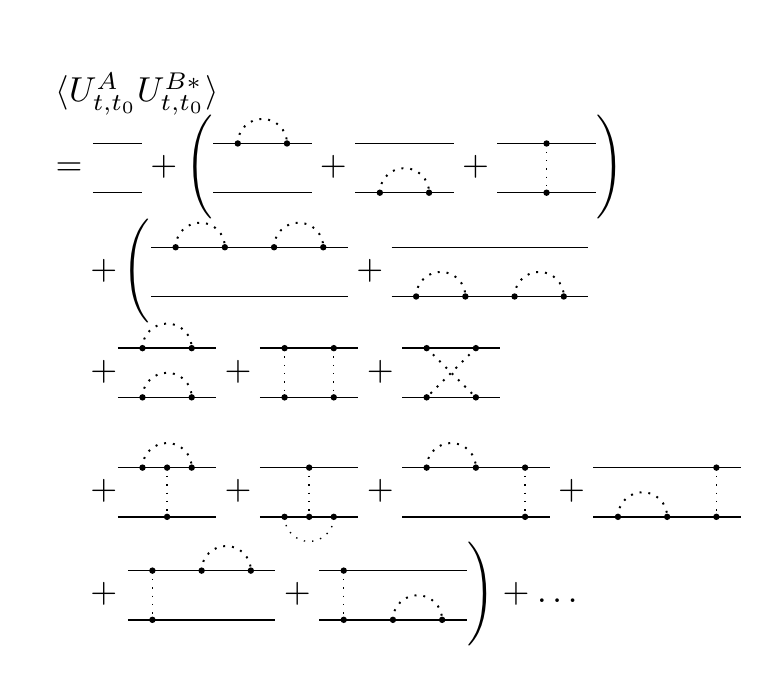}
\end{matrix}
\label{eq:diagrams}
\end{equation}
Here, the solid lines denote the ``free'' propagator \mbox{$U_{t,t_0}^{(0)} = \exp\left[-(t-t_0)\mathcal{L}\right]$} that describes particles following helical trajectories through the regular background magnetic field. The dashed lines then denote correlations between insertions of interactions with the turbulent magnetic field via the stochastic Liouville operator $\delta\mathcal{L}(t')$ at an intermediate time $t'$. All intermediate times are integrated over.

So far, we have discussed the time evolution of the angular power spectrum without specifying the exact form of $M_{\ell \ell_0}$. In fact, the mixing of angular power would happen only if $M_{\ell \ell_0}$ has non-zero off-diagonal elements. As we shall show later, one does not have mixing when considering only self-interacting diagrams in Eq.~\eqref{eq:diagrams} (diagrams with dashed lines connecting the dots on the same solid lines) since they only lead to diagonal matrices. The off-diagonal elements of $M_{\ell \ell_0}$ would actually appear when we take into account ``interaction'' diagrams (diagrams with dashed lines connecting the dots on two different solid lines). It is actually the interaction between two particles mediated by the correlation of magnetic turbulence that induces the correlation of two phase-space densities and leads to mixing of angular power from large to small scales (see Section \ref{sec:calculation-diagrams} for more details).

We could now start to evaluate the steady-state angular power spectrum for a given turbulence model if the mixing matrix for each diagrams in Eq.~\eqref{eq:diagrams} could be calculated and resummed. However, this is quite challenging even for the simplified setup considered by \citet{2019JCAP...11..048M}. Thus, we will limit ourselves to a first order calculation and only compute the diagrams up to and including the first parenthesis of Eq.~\eqref{eq:diagrams}. The difficulty in resumming the diagrams of eq.~\eqref{eq:diagrams} lies in the appearance of interaction diagrams. For the unconnected diagrams alone, the resummation of a subseries, the so-called Bourret series, is possible. This is shown in detail in appendix~\ref{sec:Bourret}. It is then possible to show that the repeated action of the time evolution operator or the mixing matrices for the unconnected diagrams is perfectly equivalent to the action of the Bourret propagator and physically represents nothing but pitch-angle scattering. We show this in appendix~\ref{sec:pitch-angle-sacttering}.

We label the contributions from the different diagrams to the correlation function as
\begin{align}
\centering
&\vcenter{\hbox{\includegraphics[scale=1,trim={0 0 0 0.4cm},clip=True]{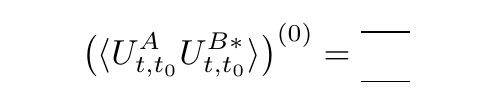}}} \\
&\vcenter{\hbox{\includegraphics[scale=1,trim={0 0 0 0.4cm},clip=True]{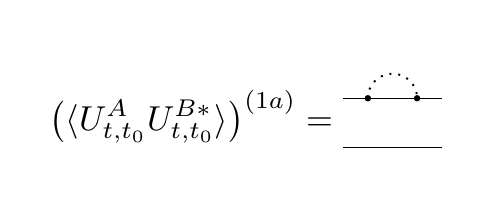}}} \label{eqn:def_U1a} \\
&\vcenter{\hbox{\includegraphics[scale=1,trim={0 0 0 0.3cm},clip=True]{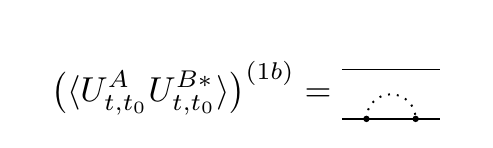}}} \\
&\vcenter{\hbox{\includegraphics[scale=1,trim={0 0 0 0.3cm},clip=True]{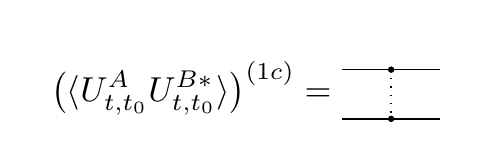}}}
\end{align}

We note again that the zeroth order diagram plus the corrections from the $(1a)$ and $(1b)$ diagrams together correspond to the usual pitch-angle scattering from QLT and are therefore not expected to lead to the mixing of the angular power spectrum. The contribution from the correlation of two phase-space densities in this formalism comes from the interaction diagram $(1c)$.

\subsubsection{The various contributions to the propagator and mixing matrix}
\label{sec:calculation-diagrams}

We note again that the time evolution operator should depend only on the time difference between the initial and final state in the case of static turbulence (see Section \ref{sec:mixing-equation}). Since we are only interested in the time evolution operator over the time step $\Delta T$, we shall from now on consider $U_{t,t_0}$ with $t=t_0+\Delta T$.

\paragraph{The (0) diagram}

Given the chosen coordinate system has the $z$-axis directed along $\vct{B}_0$, the free propagator evaluates to $U^{(0)}_{t,t_0}=\exp\left[-\ii\Omega \Delta T L_z\right]$. The double propagator then reads
\begin{equation}
\avg{\Up{A}{t}{t_0} U^{B*}_{t,t_0}}^{(0)} = \exp \left[-\ii \Omega \Delta T L^A_z + \ii \Omega \Delta T L^{B*}_z \right]
\end{equation}
which corresponds to a simple form of the leading order term for the mixing matrix element according to Eq.~\eqref{eq:mixing-matrix},
\begin{equation}
M_{\ell \ell_0}^{(0)}=\delta_{\ell\ell_0}.
\label{eq:diagram-0}
\end{equation}

\paragraph{The (1a) and (1b) diagrams}

The next-to-leading order terms (1a), (1b), and (1c) are more complicated and require a detailed discussion. In the following, we will provide the analytic formulae for these terms and their mixing matrix elements in the case of a general slab model with the turbulence tensor of the form \citep[see e.g.][and references therein]{Shalchi_book}
\begin{equation}
\avg{ \tilde{\omega}_a(\vct{k}) \tilde{\omega}^*_b(\vct{k}') } = \delta(\vct{k}-\vct{k}') g(k_\parallel) \frac{\delta(k_\perp)}{k_\parallel} \delta_{ab} \, ,
\end{equation}
if $a,b \in \{ x, y \}$ and zero otherwise. Here, $k_\parallel$ and $k_\perp$ are components of the wave vector $\vct{k}$ parallel and perpendicular to the ordered field $\vct{B}_0$. In the following we evaluate the turbulent magnetic field along  unperturbed helical CR trajectories. For a particle with momentum $\vct{p} = p \left( \sqrt{1 - \mu^2} \cos{\phi}, \sqrt{1 - \mu^2} \sin{\phi}, \mu \right)$ and position $\vct{r}_\oplus$ at time t, the position at time $t'$ reads
\begin{equation}
\begin{aligned}
\vct{r}(t') &= \vct{r}_\oplus - \uvct{x} \frac{\sqrt{1-\mu^2}}{\Omega} \sin\left[ \phi - \Omega (t-t') \right] \\
& + \uvct{y} \frac{\sqrt{1-\mu^2}}{\Omega} \cos\left[ \phi - \Omega(t-t') \right] + \uvct{z} \mu (t-t') \, .
\end{aligned}
\end{equation}

As the two terms (1a) and (1b) give the same contribution to the mixing matrix, we will illustrate the evaluation of the matrix element only for the (1a) term. The (1a) contribution to the time evolution operator is
\begin{equation}
\begin{aligned} 
& \avg{U_{t,t_0}^AU_{t,t_0}^{B*}}^{(1a)} \\
&= \!\!\int_{t_0}^t\!\!\!\mathrm{d}t_2\!\int_{t_0}^{t_2}\!\!\!\!\mathrm{d}t_1 U_{t,t_2}^{A(0)} \!\left\langle\delta\mathcal{L}_{t_2}^AU_{t_2,t_1}^{A(0)}\delta\mathcal{L}_{t_1}^A\right\rangle\! U_{t_1,t_0}^{A(0)}U_{t,t_0}^{B*(0)}\\ 
&= -\int_{0}^{\Delta T}\mathrm{d}T\int_{0}^{T}\mathrm{d}\tau \int\mathrm{d}^3\vct{k}\,g(k_\parallel)\frac{\delta(k_\perp)}{k_\perp}e^{\ii k_\parallel \mu_A \tau}\\
& \times\left[\cos{\Omega\tau}({L}^{A})^{2} + \ii \sin{\Omega\tau}{L}^A_z-\cos{\Omega\tau}({L}_z^{A})^{2}\right] \\
& \times e^{-\ii \Omega \Delta T {L}_z^A + \ii \Omega \Delta T L_z^{B*}},
\end{aligned} 
\label{eqn:U1a}
\end{equation}
where we have made the change of variables $\tau = t_2-t_1$ and $T = t_2-t_0$ for the two time integrals and also rearranged the angular momentum operators using Hadamard's lemma \citep{miller1972}.

Next, we substitute Eq.~\eqref{eqn:U1a} into Eq.~\eqref{eq:mixing-matrix} and expand the Legendre polynomials into products of spherical harmonics $Y_\ell^m(\uvct{p})$, 
\begin{equation}
P_{\ell}(\uvct{p}_A\cdot\uvct{p}_B) = \frac{4\pi}{2\ell+1}\sum_{m=-\ell}^{\ell} Y_\ell^m(\uvct{p}_A)Y_\ell^{m*}(\uvct{p}_B) \, .
\end{equation}
The integral over $\uvct{p}_B$ can be evaluated trivially and acting on the spherical harmonics with the angular momentum operators gives
\begin{equation}
\begin{aligned}
M^{(1a)}_{\ell\ell_0} &= -\int_0^{\Delta T} \mathrm{d}T\int_0^T\mathrm{d}\tau\int\mathrm{d}\vct{P}_A\int\mathrm{d}^3k \\
& \times g(k_\parallel) \frac{\delta(k_\perp)}{k_\perp} \sum_{m=-\ell}^\ell |Y_\ell^m(\uvct{p}_A)|^2\frac{e^{\ii \vct{k} \cdot \uvct{p}_A\tau}}{2\ell+1} \delta_{\ell \ell_0} \\
& \mkern-18mu \times \left[\ell(\ell+1)\cos{\Omega\tau} - \ii m \sin{\Omega\tau} - m^2 \cos{\Omega\tau} \right] \, .
\end{aligned}
\end{equation}
The sum over $m$ from the expansion of the Legendre polynomials can be evaluated using Uns\"{o}ld's theorem~ \citep{1927AnP...387..355U}, 
\begin{equation}
  \sum_{m=-\ell}^{\ell} |Y_{\ell}^{m}(\mathbf{x})|^2 = \frac{2\ell+1}{4\pi}
\end{equation} 
and its extensions,
\begin{align}
  \sum_{m=-\ell}^{\ell} m |Y_{\ell}^{m}(\mathbf{x})|^2 &= 0 \, , \\
  \sum_{m=-\ell}^{\ell} m^2 |Y_{\ell}^{m}(\mathbf{x})|^2 &= \frac{\ell(\ell+1)(2\ell+1)}{8\pi}\sin^2{\theta} \, . \label{step:unsoeld3}
\end{align}
In order to perform the remaining $\uvct{p}_A$ integration the exponential function is expanded by virtue of the plane wave expansion
\begin{equation}
  e^{i\mathbf{k}\cdot\mathbf{r}} = 4\pi\sum_{\ell=0}^\infty\sum_{m=-\ell}^{\ell}i^\ell j_\ell(kr)Y_{\ell}^{m}(\mathbf{\hat{k}})Y_{\ell}^{m*}(\mathbf{\hat{r}}),
\end{equation}
where $j_\ell(\cdot)$ denotes the spherical Bessel function of the first kind. The mixing matrix element then reads 
\begin{equation}
  \begin{aligned}
    & M_{\ell\ell_0}^{(1a)} = -\int_0^{\Delta T}\!\! \mathrm{d}T\! \int_0^T\!\! \mathrm{d}\tau\! \int \!\mathrm{d}\uvct{p}_A\! \int\mathrm{d}^3k \, g(k_\parallel)\\  
    & \times\! \frac{\delta(k_\perp)}{k_\perp}\sum_{\ell_A=0}^\infty\sum_{m_A=-\ell_A}^{\ell_A}\!\! \ell(\ell+1)\cos{\Omega\tau} \ii^{\ell_A} \\ 
    & \times j_{\ell_A}(k_\parallel\tau) Y_{\ell_A}^{m_A}(\uvct{k}) Y_{\ell_A}^{m_A*}(\uvct{p}_A) \\
    &\times
    \left(\frac{4}{3}\sqrt{\pi}Y_0^0(\uvct{p}_A)+\frac{2}{3}\sqrt{\frac{\pi}{5}}Y_2^0(\uvct{p}_A)\right) \delta_{\ell\ell_0},
    \end{aligned}
\end{equation}
where the $\uvct{p}_A$ integral can be evaluated by using the orthonormality condition of the spherical harmonics. Next, the azimuthal and perpendicular directions of the $k$ integration can be solved trivially. The $k_\parallel$ integral is rewritten as an integral over the positive part only to avoid an absolute value in the power spectrum. The $(1a)$ contribution to the mixing matrix is then given by
\begin{equation}
\begin{aligned}
M_{\ell\ell_0}^{(1a)}& = -\int_0^{\Delta T} \! \mathrm{d}T\int_0^T\mathrm{d}\tau \! \int_0^{\infty} \!\! \mathrm{d}k_\parallel g(k_\parallel)4\pi\ell(\ell+1) \\
& \times \cos{\Omega\tau}\left[\frac{2}{3}j_0(k_\parallel\tau)-\frac{1}{3}j_2(k_\parallel\tau)\right]\delta_{\ell \ell_0}.
\end{aligned}
\end{equation}
Introducing the function $\Lambda_\ell(\Delta T)$ for arbitrary integer $\ell$, 
\begin{equation}
\Lambda_{\ell}(\Delta T) \! = \!\! \int_0^{\Delta T}\!\!\!\!\!\!\! \mathrm{d}T\! \int_0^T  \mkern-11mu \mathrm{d}\tau \!\!\int_0^{\infty} \!\!\!\!\mathrm{d} k_\parallel g(k_\parallel) \cos{(\Omega\tau)} j_{\ell}(k_\parallel\tau) ,
\label{eq:diagram-1a}
\end{equation}
that will later be evaluated numerically, the mixing matrix contribution can be rewritten in the compact form 
\begin{equation}
M_{\ell\ell_0}^{(1a)} 
\! = \! - 8\pi\ell(\ell+1) \! \left[\frac{2}{3}\Lambda_0(\Delta T) - \frac{1}{3} \Lambda_2(\Delta T)\right] \! \delta_{\ell\ell_0}.
\end{equation}

\paragraph{The (1c) diagram}

The time evolution operator corresponding to the interacting diagram reads,
\begin{equation}
\label{eq:1ctimeevolution}
\begin{aligned}
&\mkern-8mu \left\langle U_{t,t_0}^A U_{t,t_0}^{B*}\right\rangle^{(1c)} \\
&=\!\!\! \int_{t_0}^t\!\!\!\mathrm{d}t_2\!\!\int_{t_0}^{t}\!\!\!\mathrm{d}t_1 U_{t,t_1}^{A(0)}
U_{t,t_2}^{B*(0)}\!\!\left\langle\delta\mathcal{L}_{t_1}^A\delta\mathcal{L}_{t_2}^{B,*}\!\right\rangle\!
U_{t_1,t_0}^{A(0)}U_{t_2,t_0}^{B*(0)}
\end{aligned}
\end{equation}

Using again Hadamard's lemma, we can simplify the operator algebra,
%
\begin{align}
e^{i\lambda L_z} L_x e^{-i\lambda L_z} &= \cos{\lambda} L_x - \sin{\lambda} L_y \, , \\
e^{i\lambda L_z} L_y e^{-i\lambda L_z} &= \sin{\lambda} L_x + \cos{\lambda} L_y \, .
\end{align}
Products of angular momentum operators can be expressed in terms of the ladder operators
\begin{align}
L_x^A L_x^B + L_y^A L_y^B &= \frac{1}{2} \left( L_{+}^A L_{-}^B + L_-^A L_{+}^B \right) \, , \\
L_x^A L_y^B - L_y^A L_x^B &= \frac{1}{2 \ii} \left( L_{-}^A L_{+}^B - L_{+}^A L_{-}^B \right) \, .
\end{align}
The action of the ladder operators on the spherical harmonics is $L_\pm Y_{\ell_0}^{m_0} = \sigma_\pm Y_{\ell_0}^{m_0\pm 1}$ and $L_\pm Y_\ell^{m_0*} = \sigma_\mp Y_{\ell_0}^{m_0\mp 1*}$ where \mbox{$\sigma_\pm = \sqrt{\ell_0(\ell_0+1)-m_0(m_0\pm1)}$}.

Using these relations the contribution from the (1c) diagram can be written as
\begin{equation}
\begin{aligned}
& \avg{ U_{t,t_0}^A U_{t,t_0}^{B*} }^{(1c)} \!\!=\!\! \int_{t_0}^{t}\!\!\mathrm{d}t_1\!\! \int_{t_0}^{t}\!\!\mathrm{d}t_2 \!\!\int\mathrm{d}^3\vct{k} \, g(k_\parallel)\frac{\delta(k_\perp)}{k_\perp} \\
& \times e^{\ii k_\parallel \mu_A(t-t_1) - \ii k_\parallel\mu_B(t-t_2)} e^{- \ii \Omega\Delta T L_z^A - \ii \Omega\Delta T L_z^B} \\
    & \times \!\bigg\{(L_x^AL_x^B\! +\! L_y^AL_y^B)\sin(\Omega(t-t_1))\sin(\Omega(t-t_2))\vphantom{\frac{}{}}\\
    & +(L_x^AL_x^B \!+\! L_y^AL_y^B)\cos(\Omega(t-t_1))\cos(\Omega(t-t_2))\vphantom{\frac{}{}}\\
    &  +(L_x^AL_y^B\! -\! L_y^AL_x^B)\cos(\Omega(t-t_1))\sin(\Omega(t-t_2))\vphantom{\frac{}{}}\\
    &  -(L_x^AL_y^B \!-\! L_y^AL_x^B)\sin(\Omega(t-t_1)\cos(\Omega(t-t_2))\vphantom{\frac{}{}}\bigg\} \, .
\end{aligned}
\end{equation}

When computing the mixing matrix, integrals over three spherical harmonics can be solved by introducing the Wigner-3j symbol
\begin{equation*}
\begin{aligned}
& \int\mathrm{d}\uvct{p} \, Y_{\ell_1}^{m_1}(\uvct{p}) Y_{\ell_2}^{m_2}(\uvct{p}) Y_{\ell_3}^{m_3}(\uvct{p}) \\
=& \sqrt{\frac{(2\ell_1\!+\!1)(2\ell_2\!+\!1)(2\ell_3\!+\!1)}{4\pi}}  \wignerj{\ell_1}{\ell_2}{\ell_3}{0}{0}{0} \!\!\! \wignerj{\ell_1}{\ell_2}{\ell_3}{m_1}{m_2}{m_3}
\end{aligned}
\end{equation*}

Combining the above equation with Eq.~\eqref{eq:mixing-matrix} the mixing matrix receives the following contribution
\begin{widetext}
\begin{equation}
\begin{aligned}
    \label{eq:diagram-1c}
    M_{\ell\ell_0}^{(1c)} =& \pi\sum_{\ell_A,\ell_B} \ii^{\ell_B-\ell_A} (2\ell_0+1) (2\ell_A+1) (2\ell_B+1) \left[ 1 + (-1)^{\ell_A+\ell_B} \right] \wignerj{\ell_A}{\ell}{\ell_0}{0}{0}{0} \wignerj{\ell_B}{\ell}{\ell_0}{0}{0}{0} \\
    & \times \sum_{m_0,m} \bigg[ (2 \ell_0 (\ell_0 + 1) - 2 m_0^2) \wignerj{\ell_A}{\ell}{\ell_0}{0}{m}{m_0} \wignerj{\ell_B}{\ell}{\ell_0}{0}{m}{m_0} \bigg] \kappa_{\ell_A,\ell_B}(\Delta T) \, ,
\end{aligned}
\end{equation}
where
\begin{equation}
  \kappa_{\ell_A,\ell_B}(\Delta T) = \int_{0}^{\Delta T}\mathrm{d}t_A\int_{0}^{\Delta T}\mathrm{d}t_B \int_0^{\infty} \mathrm{d}k_\parallel~
  g(k_{\parallel})j_{\ell_A}(k_{\parallel} t_A)j_{\ell_B}(k_{\parallel} t_B)\cos(\Omega (t_A-t_B)).
\end{equation}
\end{widetext}
Note that we have used relations presented in Appendix~\ref{appendix:1crelations} in order to shift some summation indices to achieve further simplifications. The $\kappa_{\ell_A,\ell_B}(\Delta T)$ integrals can be evaluated numerically using the Levin integration method~\citep{Levin1996FastIO}.

Due to the symmetry of exchanging particles the contribution from the $(1b)$ diagram is equal to the contribution from the $(1a)$ diagram. Since the $(0)$, $(1a)$ and $(1b)$ contributions to the mixing matrix are diagonal they do not lead to mixing between different multipoles, as expected. If it was for these two contributions only, the steady-state angular power spectrum would be only dipolar. The (1c) matrix element in Eq.~\eqref{eq:diagram-1c} is however non-diagonal and will thus lead to mixing of the dipole into higher multipoles. Putting these contributions to the mixing matrix together, Eq.~\eqref{eq:mm} can be solved numerically to get the steady state angular power spectrum.

In theory, solving Eq.~\eqref{eq:mm} requires inverting the mixing matrix for all $\ell,\ell_0$ up to infinity. In practice, however, this is not possible, but the contributions from very large $\ell,\ell_0$ on small multipoles are expected to be small. We therefore truncate the mixing matrix at $\ell_\text{max}=100$ and extrapolate the angular power spectrum as a function of $\ell_\text{max}$ to infinity assuming a power law scaling. 
We have tested the validity of this extrapolation by comparing the extrapolation to $\ell=100$ for $\ell_\text{max} = 50$ to the full calculation for $\ell_\text{max} = 100$ and found very good agreement.

\subsubsection{The free parameter $\Omega \Delta T$}
\label{sec:OmegaDT}

To determine the correct value of $\Omega\Delta T$ the resonance structure of the different contributions has to be examined. Na\"ively one would expect the relevant timescale for the durations of the wave-particle interactions to correspond to the quasi-linear scattering time $\tau_s$. This however holds only for pure pitchangle scattering. 
Starting from Eq.~\eqref{eq:1ctimeevolution} and performing the same steps as described previously for the $(1a)$ time evolution operator it can be shown that the $(1c)$ contribution in the quasi-linear theory limit becomes 
\begin{equation}
\begin{aligned}
& \avg{ U_{t,t_0}^A U_{t,t_0}^{B*} }^{(1c)} \\
&= \left(\frac{\Delta\mu}{\bar\mu}\right)^{-1}\frac{2\pi^2\Omega}{\bar\mu}g\left(\frac{\Omega}{\bar\mu}\right)\\
&\times \!\bigg[ L_+^A L_-^B \left(e^{-i\Omega\frac{\Delta\mu}{\bar\mu}\Delta T}\!\!-\!\!1\right) \!\!-\!L_-^AL_+^B \left(e^{i\Omega\frac{\Delta\mu}{\bar\mu}\Delta T}\!\!-\!\!1\right) \!\bigg]\\
&\times e^{-i\Omega L_z^A\Delta T}e^{-i\Omega L_z^{B*}\Delta T} \, ,
\end{aligned}
\end{equation}
with $\Delta \mu = \mu_A-\mu_B$ the difference of the two particles pitch angles and $\bar\mu = (\mu_A+\mu_B)/2$ the mean pitch angle. The term
\begin{equation}
\left(\frac{\Delta \mu}{\bar\mu}\right)^{-1}\left(e^{\Omega\Delta T\frac{\Delta\mu}{\bar\mu}}-1\right)
\end{equation}
has the form of a resonance with finite support for $|\Omega \Delta T \Delta\mu/\bar\mu| < 1$. Further assuming that the relative pitchangle of particles is diffusing $\langle \left(\Delta\mu/\bar\mu\right)^2\rangle \propto D_{\mu\mu}\Delta T \propto \Delta T/\tau_s$ with the pitchangle diffusion coefficient $D_{\mu\mu}$ and the quasi-linear scattering time $\tau_s$ gives
\begin{equation}
\Omega \Delta T \propto (\Omega\tau_s)^{1/3} \, .
\label{eqn:relation}
\end{equation}
The constant of proportionality depends on the magnetic turbulence geometry and the power spectrum and will be determined from numerical simulations in Sec.~\ref{sec:results_APS}.

This parameter can be interpreted as the number of gyrotimes over which correlations in the (1c) diagram decay. In QLT the particles trajectories are approximated as unperturbed trajectories. The particles can thus interact with the waves for an infinitely long time leading to a sharp resonance. In reality however particles trajectories are perturbed by the interactions with the turbulent field leading to a decay of correlations on a timescale related to the scattering time $\tau_s$.
In summary, we expect $\Omega\Delta T$ to be related to the scattering time as $\Omega\Delta T\propto (\Omega\tau_s)^{1/3}$ due to the resonance structure of the interacting diagram.

\section{Results}
\label{sec:results}

\subsection{Sky maps}
\label{sec:sky_maps}

\begin{figure*}
    \centering
    \includegraphics[scale=1]{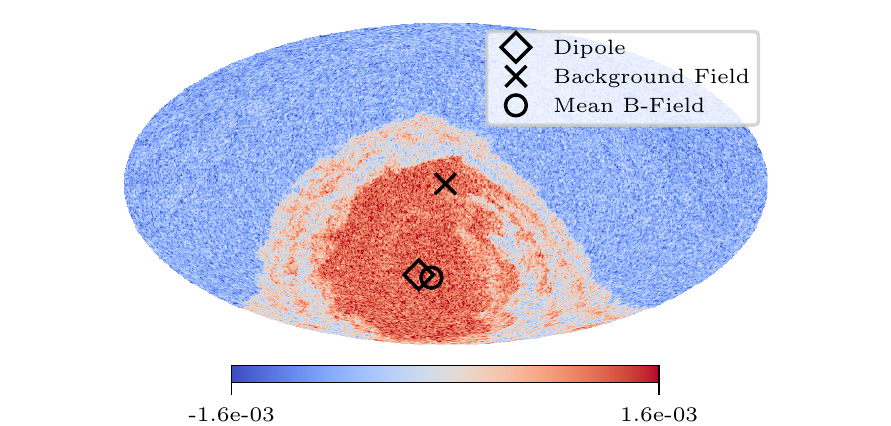}\includegraphics[scale=1]{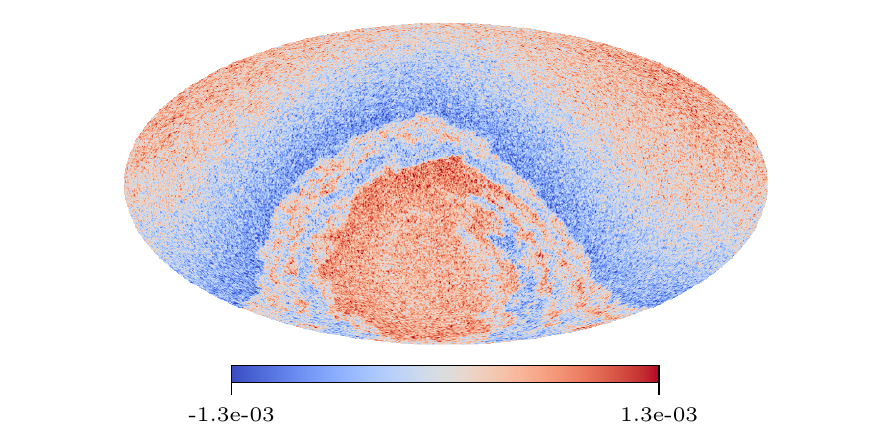}\\
    \includegraphics[scale=1]{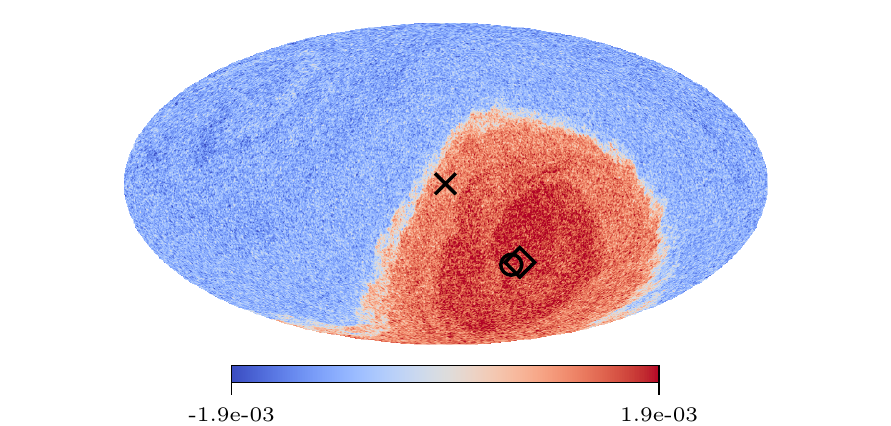}\includegraphics[scale=1]{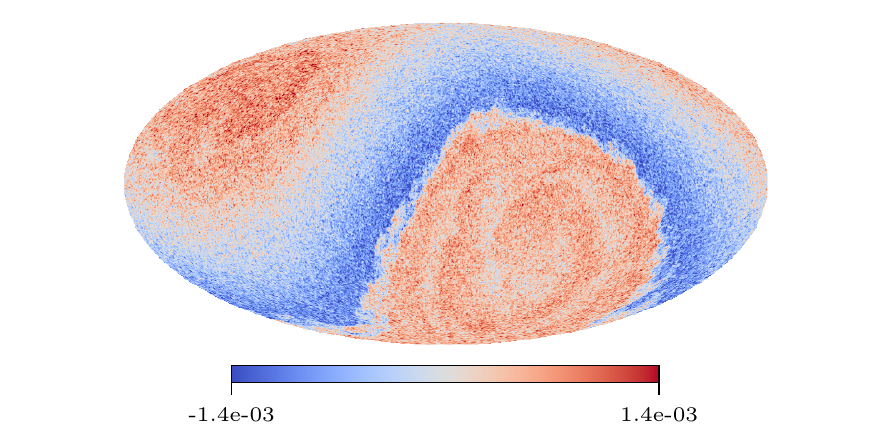}\\
    \includegraphics[scale=1]{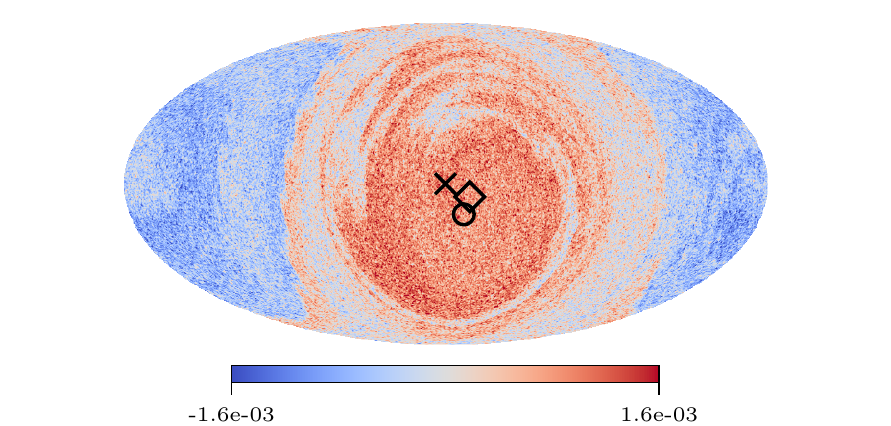}\includegraphics[scale=1]{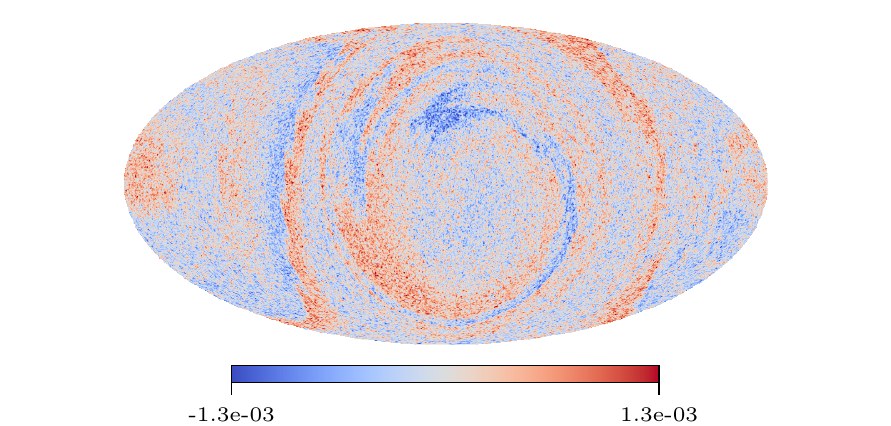}\\
    \includegraphics[scale=1]{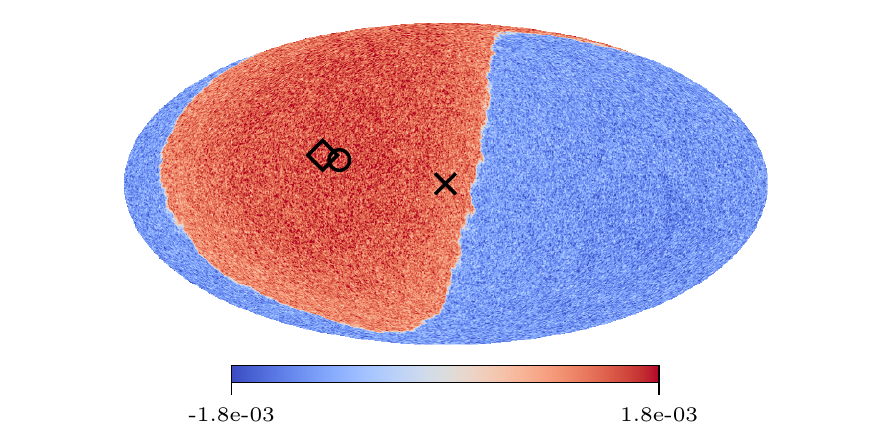}\includegraphics[scale=1]{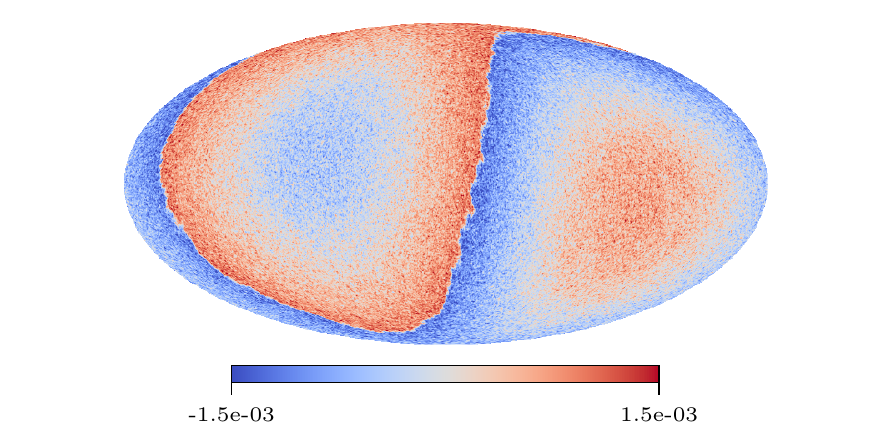}
\caption{CR anisotropy maps in various random field realizations, before (left column) and after (right column) subtraction of monopole and dipole, smoothed on a scale of $0.5^\circ$. We also compare the directions of the dipole anisotropy, the background magnetic field and the mean magnetic field averaged over a length of $L_\text{avg} = (1/30) L_c$. The simulations are done at turbulence level $\eta = \delta B^2/(B_0^2+\delta B^2) = 0.5$ and at $\rho=r_g/L_c = 8.9\times 10^{-4}$. We adopt a backtracking time $\Omega T = 2000$.}
    \label{fig:dipoledir}
\end{figure*}

We present sky maps of the CR arrival directions in four different random realizations of the turbulent magnetic field in the left column of Fig.~\ref{fig:dipoledir}. These sky maps are chosen at $T \gtrsim \tau_s$ since we do not expect significant variations of the large-scale features in these sky maps (except for the contamination by the noise, see Sec.~\ref{sec:results_APS}) once particles are backtracked into the diffusive regime. This might be better visualized in Fig.~\ref{fig:illustration} where the shape of the CR excess region (the red patch in the sky map) remains roughly unchanged as we proceed to larger backtracking time. Roughly speaking, this means that the angular power spectrum on large scales converges for $T\gtrsim \tau_s$ as we will discuss in depth in Sec.~\ref{sec:results_APS}. We also see that a dipole appears in all realizations as the dominant anisotropy. Interestingly, the direction of the dipole vector in each realization $i$ (diamonds),
\begin{eqnarray}
\vct{\Phi}_i=\frac{1}{4\pi}\int\textrm{d}\uvct{p}\,\uvct{p}\frac{f(\vct{r}_\odot,\uvct{p},t)}{\bar{f}}
\end{eqnarray}
seems to point in the direction of the local mean magnetic field (empty circles) and not in the direction of the regular background magnetic field (crosses). 

The right column of Fig.~\ref{fig:dipoledir} shows the sky maps for the same magnetic field realizations as in the left column of Fig.~\ref{fig:dipoledir} with the monopole and dipole fitted and subtracted. The sky maps still clearly show signs of anisotropy.

\begin{figure}
    \centering
    \includegraphics[scale=1]{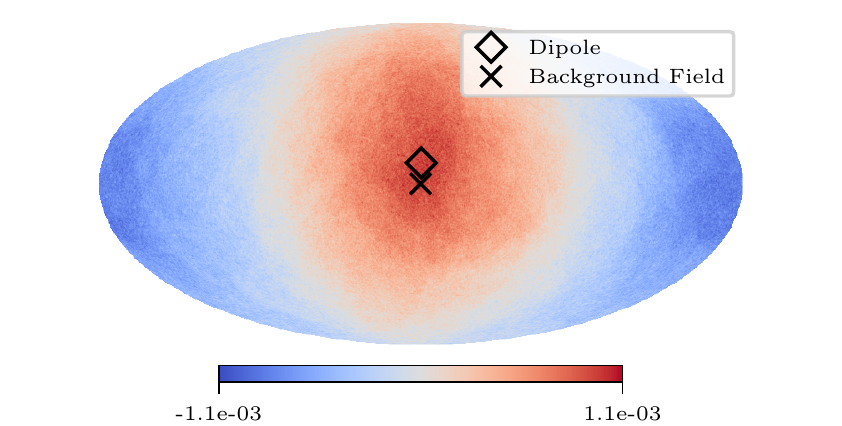} \includegraphics[scale=1]{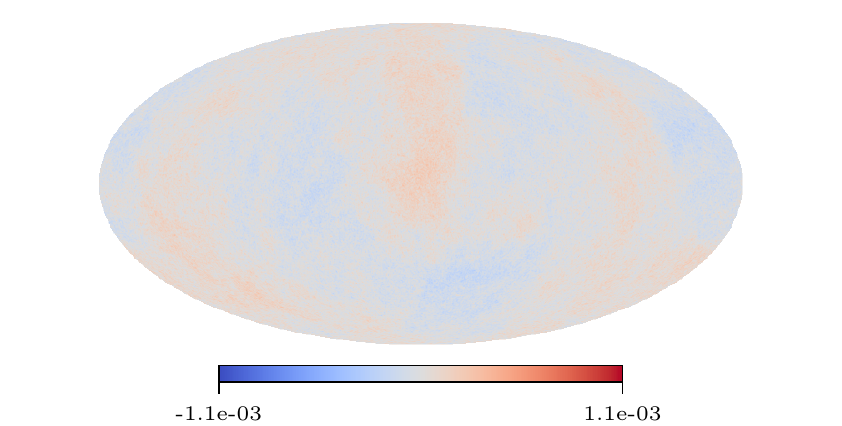}\\ 
    \caption{Ensemble averaged sky map of the CR anisotropy from test particle simulations in slab turbulence with turbulence level $\eta = 0.5$ at normalized rigidity $\rho = r_g/L_c \simeq 8.9\times 10^{-4}$ averaged over 35 magnetic field realizations. The bottom panel shows the same ensemble averaged sky map with the monopole and dipole subtracted.}
    \label{fig:skymaps_qlt}
\end{figure}

The results of QLT can be approximated by averaging the phase-space density over the ensemble of magnetic field realizations.
The ensemble averaged phase-space density calculated from the test particle simulations by averaging over $N=35$ different field realizations is shown in Fig.~\ref{fig:skymaps_qlt}. As expected from QLT, the dipole component of the averaged map,
\begin{eqnarray}
\langle\vct{\Phi}\rangle = \frac{1}{N}\sum_{i=1}^{N}\vct{\Phi}_i,
\end{eqnarray}
is pointing in the direction of the regular magnetic field. The misalignment between $\langle \vct{\Phi} \rangle$ and $\vct{\Phi}_i$ in all realizations is due to a relatively large value of $\eta$ adopted for these simulations and, as we shall see later in Sec.~\ref{sec:dipole}, this effect has to be taken into account in order to reconcile the normalization of the numerical and analytic angular power spectra for strong turbulence. We show also the ensemble averaged phase-space density with the monopole and dipole removed in the bottom panel of Fig.~\ref{fig:skymaps_qlt}. It has significantly less structure than the corresponding sky maps in the right column of Fig.~\ref{fig:dipoledir}, but a small octupole is visible. 

\subsection{Angular power spectrum}
\label{sec:results_APS}

\begin{figure}
    \centering
    \includegraphics[scale=1]{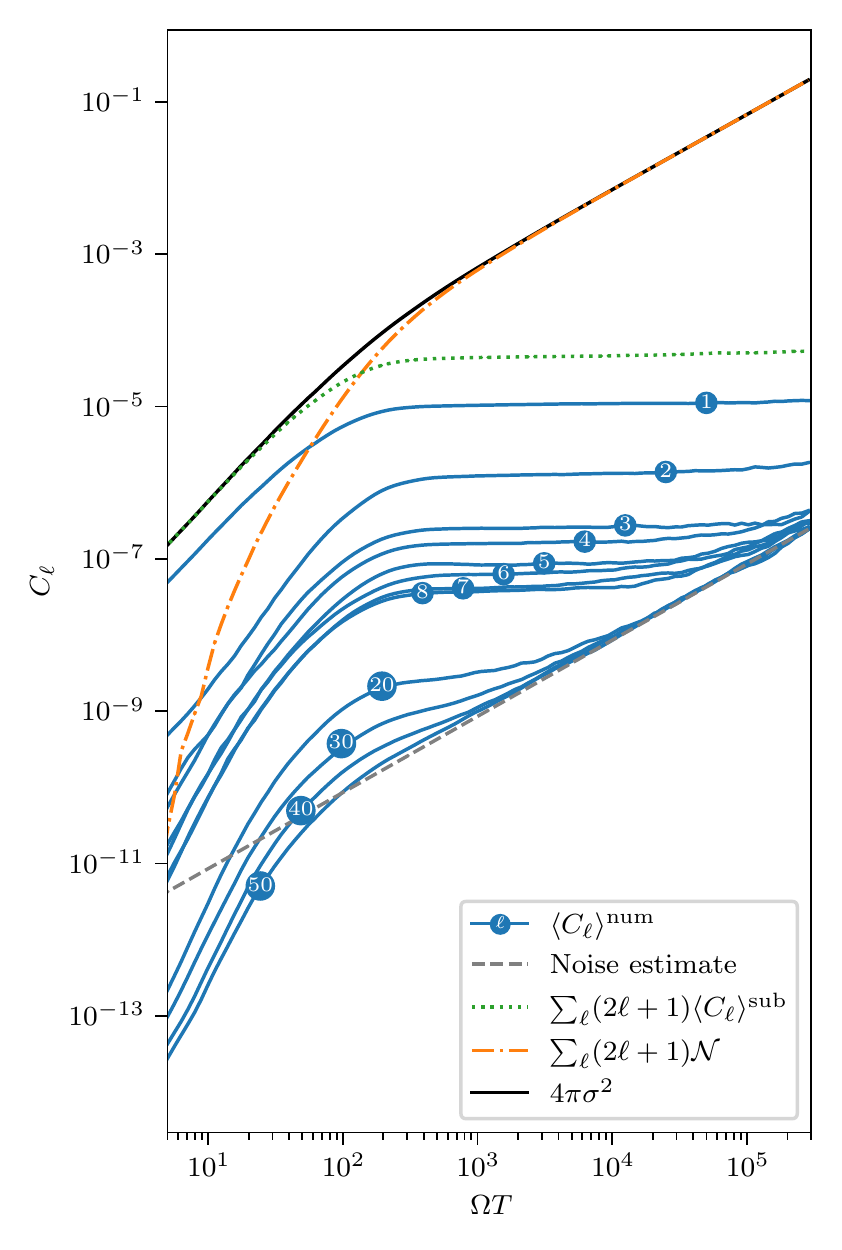}
    \caption{Ensemble-averaged angular power spectrum as a function of backtracking time $T$ from the test particle simulations for a normalized rigidity of $\rho \simeq 8.9\times 10^{-3}$. The solid lines show the different multipoles, the dashed line shows the estimated noise level growing linearly with time. The other lines show the total noise subtracted angular power (green dotted line), the noise total angular power (orange dot-dashed line) and the sky map total angular power (solid black line).}
    \label{fig:time-dependent}
\end{figure}

Fig.~\ref{fig:time-dependent} shows the ensemble-averaged angular power spectrum from simulations $\langle C_\ell\rangle^{\rm num}$ as a function of backtracking time $T$ for the normalised rigidity \mbox{$\rho = r_{\text{g}} / L_{\text{c}} \simeq 8.9\times 10^{-3} $}, which corresponds to $R = 1 \, \text{PV}$ for the parameters detailed in Sec.~\ref{sec:parameters}. All the multipoles initially grow with $T$. For $T \gtrsim \tau_s$, the angular power spectrum on large scales starts converging towards a constant value (for the adopted parameters of turbulence $\tau_s(R=1\textrm{ PV})\simeq 100\,\Omega^{-1}$). We can also see that all the multipoles, especially the ones on small scales, are distorted by the shot noise at sufficiently large values of $T$ since $\mathcal{N}_\ell\sim T$ for $T\gtrsim \tau_s$, see Eq.~\eqref{eqn:noise}. In fact, the ensemble-averaged noise-subtracted angular power spectrum $\langle C_\ell\rangle^{\rm sub}=\langle C_\ell\rangle^{\rm num}-\mathcal{N}_\ell$ is expected to become constant for $T\gtrsim \tau_s$ which is essentially due to the angular power spectrum being sensitive only to the particular realization of the \emph{local} magnetic field. Clusters of particles being tracked back through time away from the observer initially travel ballistically and get deflected coherently as they experience similar magnetic fields. When they reach the regime of diffusive transport the trajectories decorrelate and hence the sky maps becomes noisier, but the ensemble-averaged, noise-subtracted angular power spectrum stays constant. The angular power spectrum is thus the imprint of the turbulent magnetic field in the transitory regime between ballistic and diffusive particle transport.

In addition, we present the noise-subtracted total angular power, which is $\sum_{\ell}(2\ell+1)\langle C_\ell\rangle^{\rm sub}$, as the green dotted line in Fig.~\ref{fig:time-dependent}. Note that in the summation we have used the empirical relation \mbox{$\langle C_\ell\rangle^{\rm sub}\sim \ell^{-2.7}$}, to be shown below, for $\ell\geq 5$. The sky-map total angular power, that is the ensemble-averaged variance of the sky maps $4\pi\sigma^2=\sum_{\ell}(2\ell+1)\langle C_\ell\rangle^{\rm num}$, is also shown as the solid black line which grows linearly with $T$ for $T\gg \tau_s$, see Eq.~\eqref{eqn:sky_variance}. More importantly, we have found that $4\pi\sigma^2\gg \sum_{\ell}(2\ell+1)\langle C_\ell\rangle^{\rm sub}$ for $T\gg\tau_s$ and, thus, the dominant contribution to the sky-map total angular power at large backtracking time is the noise total angular power $\sum_{\ell}(2\ell+1)\mathcal{N}_\ell$ (orange dot-dashed line in Fig.~\ref{fig:time-dependent}).

We note also that this result is valid regardless of the number of pixels $N_{\rm pix}$ of the simulations since 
the sum over $\ell$ with an $N_\text{pix}$ dependent summation range exactly cancels the $N_\text{pix}$ dependence of $\mathcal{N}_\ell$, making the noise total angular power independent of $N_\text{pix}$.

\begin{figure*}[tbh]
  \centering
  \includegraphics[width=\textwidth]{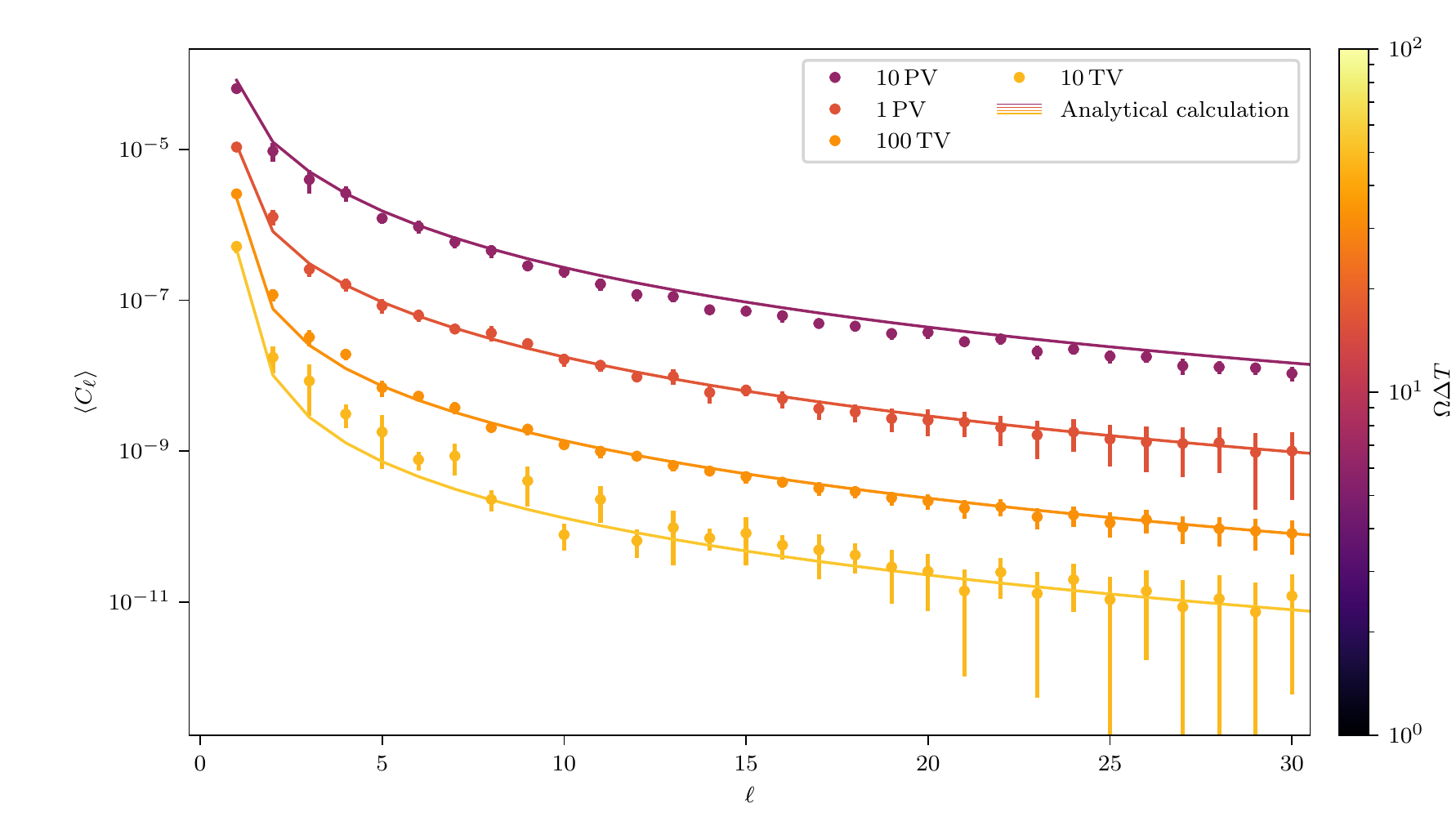}
  \caption{Angular power spectra of the arrival directions of CRs. The points show the angular power spectra computed from the results of test particle simulations in slab turbulence with a turbulence level of $\eta = 0.5$. Also shown is the best fit line from the analytical results for each individual rigidity. The numerical and analytical results agree well down to the smallest angular scales.}
  \label{fig:fit}
\end{figure*}

The noise-subtracted angular power spectra from the numerical simulations, normalised to the dipole, for the normalised rigidities $\rho = r_{\text{g}} / L_{\text{c}} \simeq 8.9 \times 10^{-5}$, $8.9\times 10^{-4}$, $8.9\times 10^{-3}$ and $8.9\times 10^{-2}$ are shown in Fig.~\ref{fig:fit}.  With the parameters as adopted in Sec.~\ref{sec:parameters}, that is $B_\text{RMS} = 4\,\mu\text{G}$ and $L_c = 30\,\text{pc}$, this corresponds to the rigidities $10 \, \text{TV}, 100 \, \text{TV}, 1 \, \text{PV}$ and $10 \, \text{PV}$. The angular power spectra are falling power laws in $\ell$ for $\ell \gtrsim 2$ and the magnitude of the spectral index is larger for smaller rigidities.

\begin{figure}[tbh]
    \centering
    \includegraphics[scale=1]{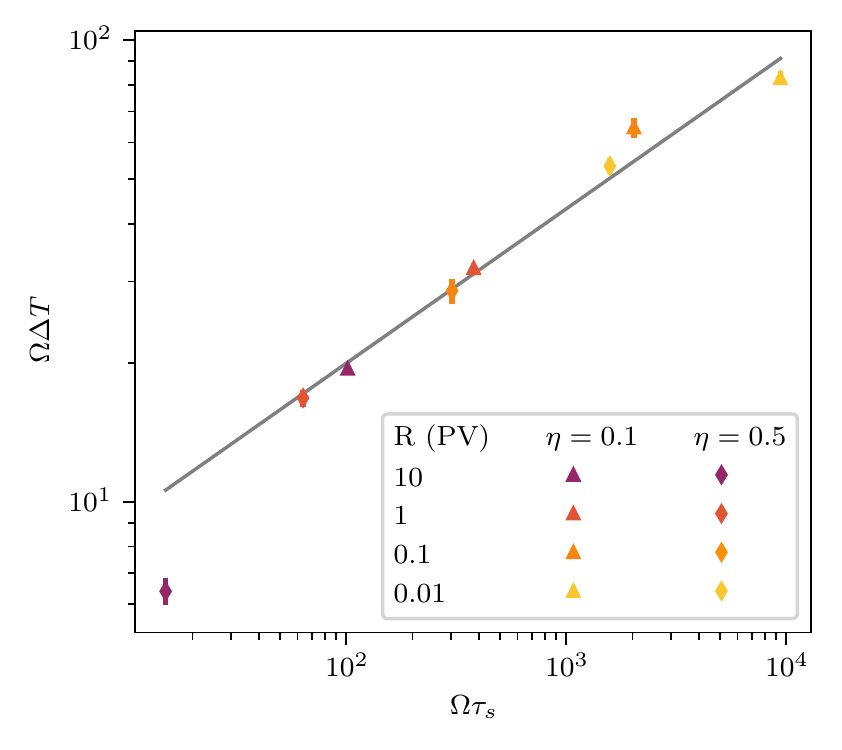}
    \caption{Best fit $\Omega \Delta T$ parameter as a function of the quasi-linear scattering time $\Omega \Delta \tau_s$ for simulations at different rigidities and different turbulence levels. The solid line shows the fitted relation $\Omega \Delta T \simeq 4 (\Omega\tau_s)^{1/3}$.}
    \label{fig:taus}
\end{figure}

Also shown in Fig.~\ref{fig:fit} are the steady-state angular power spectra from the analytical computation, showing excellent agreement with the numerical results. We have also normalised to the dipole here, but we discuss this normalisation in Sec.~\ref{sec:dipole} below. In computing the mixing matrices, we have adopted the relation between the free parameter $\Omega \Delta T$ and the quasi-linear scattering time $\Omega \tau_{\text{s}}$ predicted in Eq.~\eqref{eqn:relation}. This is a prediction up to the constant of proportionality and we have found this constant to be $\simeq 4$ by fitting to the results from the numerical simulations, that is $\Omega \Delta T \simeq 4 (\Omega\tau_s)^{1/3}$. This fit is illustrated in Fig.~\ref{fig:taus}. The steady state angular power spectra initially fall off very quickly depending on rigidity via the $\Omega\Delta T$ parameter and then, for $\ell\gtrsim 5$ exhibit power-law like falling behaviour with a spectral index close to $-2.7$ for all rigidities. This results in a slightly flatter angular power spectrum than the $\ell^{-3}$ scaling that was found in~\cite{2014PhRvL.112b1101A} and is more similar to the scaling found in~\cite{2015ApJ...815L...2A} for their model parameter $p=0.6$.

\subsection{Dipole amplitude}
\label{sec:dipole}

We have already shown that the dipole typically aligns with the average magnetic field direction instead of the background field direction, see Sec.~\ref{sec:sky_maps}. It turns out that the normalisations of the angular power spectra from the numerical simulations and from the analytical calculation differ, while the shapes agree. We will discuss the origin of this disagreement and explain how to correct the overall normalisation of the angular power spectra from the analytical calculation.

Both of these results could be better understood by looking at the dipole direction within each realization which can be related to the dipole strength as follows
\begin{eqnarray}
C^{(i)}_1=4\pi\vct{\Phi}_i^2.
\end{eqnarray}
This quantity is technically the ``observed'' dipole strength for a specific realization of turbulence and, correspondingly, the numerical or noise subtracted ensemble-averaged dipole strength (before the noise also starts dominating the dipole) is 
\begin{eqnarray}
\langle C_1 \rangle^{\rm sub}\simeq \langle C_1 \rangle^{\rm num}=\frac{4\pi}{N}\sum_{i=1}^{N}\vct{\Phi}_i^2=\frac{1}{N}\sum_{i=1}^{N}C_1^{(i)}.\label{eq:C1-observe}
\end{eqnarray}
where $N$ is the total number of realizations considered. Note again that this statement is only true if the dipole is not dominated by the shot noise which should be the case for $T$ roughly a few times $\tau_s$. In the standard QLT approach, we could also predict the ensemble average of the dipole vector following Fick's law \citep{2017PrPNP..94..184A}
\begin{eqnarray}
\langle\vct{\Phi}\rangle \simeq \frac{1}{N}\sum_{i=1}^{N}\vct{\Phi}_i=-\frac{\vct{K}\cdot\nabla\bar{f}}{\bar{f}},
\label{eq:dipole}
\end{eqnarray}
from which we could also write down the QLT ensemble average dipole strength $\langle C_1 \rangle^{\rm QLT}$ as
\begin{eqnarray}
\langle C_1 \rangle^{\rm QLT} = 4\pi \langle\vct{\Phi}\rangle^2 = \frac{4\pi}{N^2} \sum_{i,j=1}^{N} \vct{\Phi}_i \vct{\Phi}_j.\label{eq:C1-QLT}
\end{eqnarray}

Having presented the various definitions of the dipole, we can now compare them with each other. In the following, we shall fix the direction of the CR gradient along $\vct{B}_0$ for simplicity. This means that the ensemble average dipole direction should be aligned with $\vct{B}_0$ as could be seen from Eq. \eqref{eq:dipole}. This might not be too problematic for a small turbulence level as we expect the dipoles in all realizations to point more or less in the direction of $\vct{B}_{0}$ which, roughly speaking, means $\vct{\Phi}_i\simeq\langle \vct{\Phi} \rangle$ or equivalently $\langle C_1\rangle^{\rm sub} \simeq \langle C_1\rangle^{\rm QLT}$. If the turbulence level is large enough, e.g. $\langle\delta\vct{B}^2\rangle\simeq\vct{B}_0^2$, the prediction for the dipole might not be accessible in a straightforward way within the framework of QLT \citep{mertsch2015}. This is because CRs propagate principally along the large-scale magnetic field $\vct{B}_i$ within the $i$th realization meaning the flux of particles determining the dipole direction should also point along $\vct{B}_i$ which has a large contribution from the turbulent field and do not exactly follow $\vct{B}_0$. More importantly, Eq. \eqref{eq:C1-observe} and \eqref{eq:C1-QLT} also tell us that $\langle C_1\rangle^{\rm sub} > \langle C_1\rangle^{\rm QLT}$ for strong turbulence (see also \cite{2015ApJ...815L...2A} for the case of isotropic turbulence). We shall now explore this scenario with test particle simulations. 

The sky maps of CR arrival directions within a few different realizations of the turbulence with $\eta=0.5$ were already shown in Fig. \ref{fig:dipoledir}. Note that the direction of the $\vct{B}_0$-field is in the middle of the map. The direction of the local large-scale magnetic field $\vct{B}_i$ within each realizations (circle) could also be obtained by performing the spatial average of the turbulent field over $L_{\rm avg}$. We note that $L_{\rm avg}$ has to be sufficiently smaller than the correlation length. The results are shown for $L_{\rm avg}=L_c/30$, but they are insensitive to the exact choice of this parameter.

\begin{figure}
    \centering
    \includegraphics[scale=1]{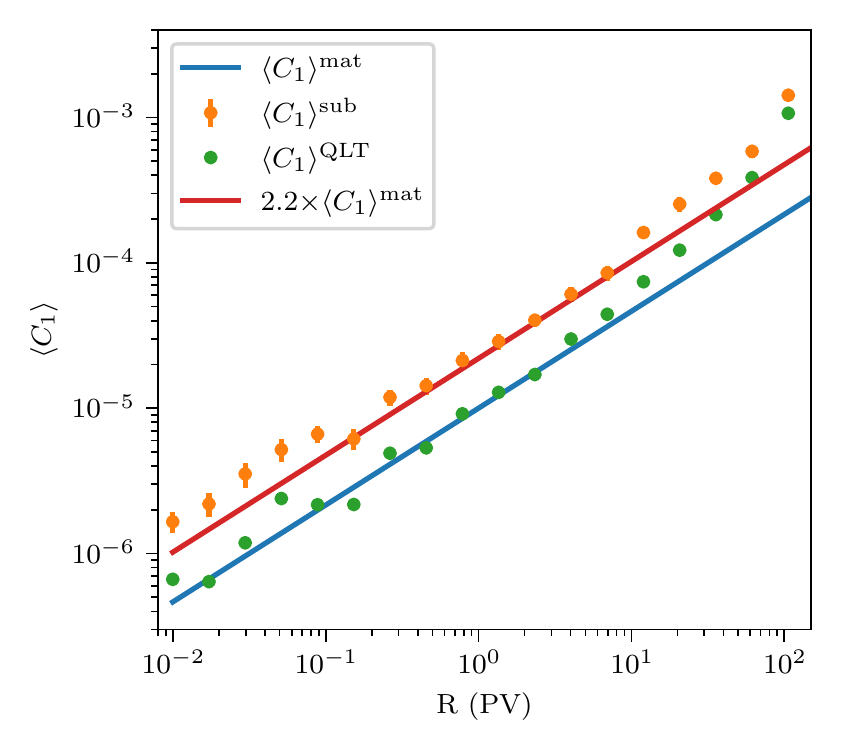}
    \caption{Noise subtracted and QLT dipole strengths at different rigidities for test particle simulations with a turbulence level $\eta=0.5$ in comparison to the mixing matrix prediction.}
    \label{fig:dipole-amplitude}
\end{figure}

We also present the dipole strength for different rigidities in Fig.~\ref{fig:dipole-amplitude}. The green and orange points represent respectively the ensemble-averaged QLT and noise-subtracted dipole strength obtained from the numerical simulations. We also include the dipole strengths at different rigidities from the mixing matrix approach $\langle C_\ell\rangle^{\rm mat}$ (solid blue line) and they seem to match very well with $\langle C_1\rangle^{\rm QLT}$. This is understandable in the sense that the mixing matrix formalism only considers unperturbed trajectories and particles propagating along $\vct{B}_0$ similar to the standard QLT approach as indicated in Eq. \eqref{eq:product-small-time} where we have assumed $\lim_{\Delta T\rightarrow 0}\langle \Delta \vct{r}\rangle =\mu\Delta T\uvct{z}$. 

It is also clear that $\langle C_1\rangle^{\rm sub} >  \langle C_1\rangle^{\rm mat} \simeq \langle C_1\rangle^{\rm QLT}$ since we are in the case of $\eta=0.5$. Note that we have assumed $\nabla\bar{f}$ to follow the same scaling in rigidity as $\bar{f}$ and, thus, $\langle C_\ell\rangle^{\rm QLT}\sim K_{zz}^2 \sim R^{2/3}$. Interestingly, $\langle C_\ell\rangle^{\rm sub}$ seems to have the same rigidity scaling as $\langle C_\ell\rangle^{\rm QLT}$ with the normalization shifted by a factor of about two. Näively, we might expect $\vct{B}_i$ and $\vct{B}_0$ to make an angle of around $\alpha\simeq\pi/4$ in the case of $\eta=0.5$ which means the projection of the dipole vectors $\vct{\Phi}_i$ along $\vct{B}_0$ would make $\langle C_1\rangle^{\rm sub}/\langle C_1\rangle^{\rm mat} \simeq 1/\eta \simeq 1/\cos^2\alpha \simeq 2$. 

In principle, we should be able to predict $\langle C_\ell\rangle^{\rm mat}$ more accurately and better match $\langle C_\ell\rangle^{\rm sub}$ even for strong turbulence once the particle's motion along $\vct{B}_i$ is taken into account. This requires a description of the CR transport perpendicular to $\vct{B}_0$. However, we expect this effect to modify only the dipole source term in Eq. \eqref{eq:mm} by introducing an additional factor which depends only on the turbulence level $\eta$ and, thus, the relation between $\tau_s$ and $\Delta T$ obtained from fitting the shape of the angular power spectrum spectra should remain roughly the same. For now, we shall adopt a more straightforward phenomenological approach which is to obtain the cosine of the misalignment angle $\alpha$ between $\vct{B}_i$ and $\vct{B}_0$ for all realizations and then rescale $\langle C_\ell\rangle^{\rm mat}$ by a factor $1/\langle\cos\alpha\rangle^2\simeq 2.2$.

In addition, due to limited dynamical range in the numerical testparticle simulations the diffusion coefficient at the smallest rigidities deviates from the expected $R^{-1/3}$ scaling and thus the normalization of the angular power spectrum is sligthly too large. Therefore the numerical angular power spectrum at $10\,\text{PV}$ has to be rescaled to give the correct rigidity dependence of the dipole anisotropy.
This provides good fits of $\langle C_\ell\rangle^{\rm mat}$ to $\langle C_\ell\rangle^{\rm sub}$ as shown in Fig.~\ref{fig:fit}. 

\section{Summary and conclusion}
\label{sec:summary}

The angular power spectrum of CR arrival directions is an important observable. As it is sourced by correlations of particles experiencing the same turbulent magnetic field it can be used to infer the local field configuration. Understanding the origin of the angular power spectrum is therefore important as an independent probe of the outer scale of turbulence and the local turbulence geometry.

Here we have used a perturbative calculation to predict the angular power spectrum of CR arrival directions taking into account the correlations of phase-space densities implied by the correlations in the turbulent magnetic field. We have assumed a homogeneous background magnetic field such that the unperturbed orbits are helical. The perturbative expansion up to first order in the turbulence strength then includes resonance effects between particles and the turbulent magnetic field similarly to QLT. The difference in the angular power spectrum compared to QLT arises because also correlations between phase-space densities that are induced by particles travelling through the same field realization are treated explicitly. This leads to a finite angular power spectrum even at larger multipoles $\ell$.

To validate and test the assumptions that were made in this calculation we have compared to test particle simulations done in the same turbulence model at the rigidities relevant for observations by IceCube and HAWC. This comparison shows very good agreement between the analytical model and the numerical test particle simulations. 

If the turbulence in the local interstellar medium indeed has the slab geometry, we can use this model to constrain the parameters of the turbulent magnetic field. In particular, fitting the shape of the angular power spectrum to the available observational data will allow us to probe the scattering time which might, in turn, provide more information on the correlation length and the turbulence level in our Galactic neighborhood. This will however require more information on the uncertainties from observations. It is clear that the analytic model always predicts the angular power spectrum to be rigidity-dependent and yet sky maps of the arrival directions are observed for particles of different energies. In fact, the conversion between rigidity and energy might not be straightforward as the CR compositions assumed could vary between observatories. More importantly, even if the CR composition is known with confidence, we have to note that observational data always come with a finite energy resolution and, thus, the observed anisotropies might contain features different from the ansiotropies expected for individual energies. Finally, we should note also that the limited sky coverage of observatories like IceCube and HAWC could also hinder the reconstructions of the angular power spectrum from observed sky maps. Futher investigations are needed in order to better address these problems and gain more insight into the nature of local turbulence hidden in the small-scale anisotropies.   

\begin{acknowledgments}
This project was funded by the Deutsche Forschungsgemeinschaft (DFG, German Research Foundation) -- project number 426614101. 
\end{acknowledgments}

\bibliographystyle{aasjournal}
\bibliography{Bibliography}

\appendix
\section{Bourret series}
\label{sec:Bourret}

In Sec.~\ref{sec:mixing_matrix}, we have expanded the ensemble-averaged product of two propagators into a diagrammatic series. This can also be done for the ensemble-average of a single propagator and the ensuing series, the so-called Bourret series, can be resummed.

The sum of all connected diagrams defines the so-called mass operator,
\begin{equation}
\vcenter{\hbox{
\includegraphics[scale=1,trim={0.7cm 0.5cm 0.8cm 0.7cm}, clip=false]{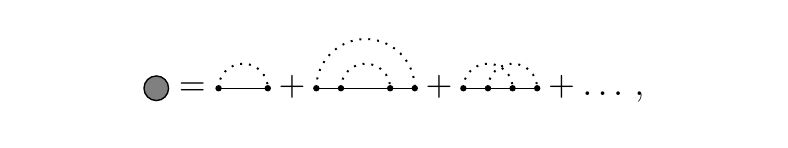}
}}
\end{equation}
and so the full propagator reads
\begin{equation}
\vcenter{\hbox{
\includegraphics[scale=1,trim={0.2cm 0.7cm 0.1cm 0.8cm}, clip=false]{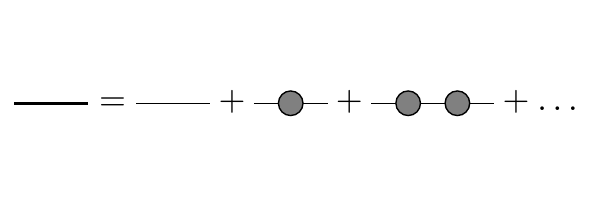}
}}
\end{equation}
Given the complicated nestings of $2$-point functions, the mass operator will be difficult to evaluate. However, approximating the mass operator with its lowest order term, we obtain the so-called Bourret series,
\begin{equation}
\vcenter{\hbox{
\includegraphics[scale=1,trim={0 0.8cm 0 0.7cm}, clip=false]{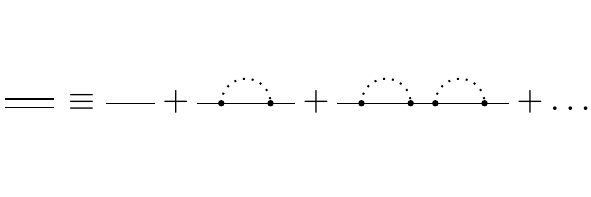}
}}
\label{eqn:Bourret_single1}
\end{equation}

It is straight-forward to resum the Bourret series, eq.~\eqref{eqn:Bourret_single1}, after Laplace transformation with respect to $(t-t_0)$. Courtesy of the convolution theorem, the nested convolutions in intermediate times transform to products. We illustrate this with the example of the second correction to the free propagator,
\begin{align}
& \int_0^{\infty} \dd (t-t_0) \, \ee^{-s(t-t_0)} \left( \vcenter{\hbox{
\includegraphics[scale=1,trim={0.5cm 0 0.5cm 0}, clip=true]{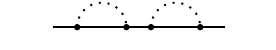}
}} \right) \\
&= \int_{-\infty}^t \dd t_0 \, \ee^{-s(t-t_0)} \left( \vcenter{\hbox{
\includegraphics[scale=1,trim={0.5cm 0 0.5cm 0}, clip=true]{figures/single1_2nd_order}
}} \right) \\
&= \int_{-\infty}^t \dd t_4 \int_{-\infty}^{t_4} \dd t_3 \int_{-\infty}^{t_3} \dd t_2 \int_{-\infty}^{t_2} \dd t_1 \int_{-\infty}^{t_1} \dd t_0 \ee^{-s(t-t_0)} U_{t,t_4}^{(0)} \left\langle \delta\mathcal{L}(t_4) U_{t_4,t_3}^{(0)} \delta\mathcal{L}(t_3) \right\rangle \\
& \qquad \times U_{t_3,t_2}^{(0)} \left\langle \delta\mathcal{L}(t_2) U_{t_2,t_1}^{(0)} \delta\mathcal{L}(t_1) \right\rangle U_{t_1,t_0}^{(0)} \\
&= \left( \int_{-\infty}^t \dd t_4 \ee^{-s(t-t_4)} U_{t,t_4}^{(0)} \right) 
\left( \int_{-\infty}^{t_4} \dd t_3 \, \ee^{-s(t_4-t_3)} \left\langle \delta\mathcal{L}_{t_4} U_{t_4,t_3}^{(0)} \delta\mathcal{L}_{t_3} \right\rangle \right) \\
& \times \left( \int_{-\infty}^{t_3} \dd t_2 \ee^{-s(t_3-t_2)} U_{t_3,t_2}^{(0)} \right)
\left( \int_{-\infty}^{t_2} \dd t_1 \, \ee^{-s(t_2-t_1)} \left\langle \delta\mathcal{L}_{t_2} U_{t_2,t_1}^{(0)} \delta\mathcal{L}_{t_1} \right\rangle \right) \\
& \times \left( \int_{-\infty}^{t_1} \dd t_0 \ee^{-s(t_1-t_0)} U_{t,t_0}^{(0)} \right) \\
&= \tilde{U}(s) \tilde{V}(s) \tilde{U}(s) \tilde{V}(s) \tilde{U}(s) \, . \label{eqn:second_correction-Laplace-transfd}
\end{align}
Here, we have defined the Laplace transforms of the free propagator, $\tilde{U}(s)$ and of the Bourret insertion, $\tilde{V}(s)$,
\begin{align}
\tilde{U}(s) & \equiv \int_{-\infty}^t \dd t' \ee^{-s(t-t')} U_{t,t'}^{(0)} \, , \\
\tilde{V}(s) & \equiv \int_{-\infty}^t \dd t' \, \ee^{-s(t-t')} \left\langle \delta\mathcal{L}_t U_{t,t'}^{(0)} \delta\mathcal{L}_{t'} \right\rangle \, .
\end{align}

It is not difficult to see that the Laplace transform of the Bourret series, eq.~\eqref{eqn:Bourret_single1}, thus reduces to the geometric series,
\begin{equation}
\int_{-\infty}^t \dd t_0 \, \ee^{-s(t-t_0)} \left( \vcenter{\hbox{
\includegraphics[scale=1,trim={0.5cm 0 0.5cm 0},clip=true]{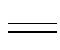}
}} \right) = \tilde{U}(s) \sum_{n=0}^{\infty} \left( \tilde{U}(s) \right)^n \left( \tilde{V}(s) \right)^n = \frac{\tilde{U}(s)}{1 - \tilde{U}(s) \tilde{V}(s)} \, . \label{eqn:Laplace_transform_Bourret_series}
\end{equation}

The Laplace transform of the free propagator, $\tilde{U}(s)$ evaluates to
\begin{equation}
\tilde{U}(s) = \frac{1}{s - \ii \Omega L_z} \, .
\end{equation}
And so performing the inverse Laplace transform of eq.~\eqref{eqn:Laplace_transform_Bourret_series} gives
\begin{align}
\vcenter{\hbox{
\includegraphics[scale=1]{figures/single_double_line}
}} &= \frac{1}{2 \pi \ii} \int_{\gamma - \ii \infty}^{\gamma + \ii \infty} \dd s \, \ee^{s(t-t_0)} \frac{1}{s - \ii \Omega L_z - \tilde{V}(s)} \\
&\simeq \frac{1}{2 \pi \ii} \int_{\gamma - \ii \infty}^{\gamma + \ii \infty} \dd s \, \ee^{s(t-t_0)} \frac{1}{s - \ii \Omega L_z - \tilde{V}(0)} \\
&= \ee^{(\ii \Omega L_z + \tilde{V}(0)) (t-t_0)} \, ,
\end{align}
where we have anticipated the limit $(t - t_0) \to \infty$ in order to approximate $\tilde{V}(s) \simeq \tilde{V}(0)$.

It thus remains to compute $\tilde{V}(0)$,
\begin{align}
\tilde{V}(0) &= \int_{-\infty}^t \dd t' \, \left\langle \delta\mathcal{L}_t U_{t,t'}^{(0)} \delta\mathcal{L}_{t'} \right\rangle \\
&= - \int_{-\infty}^t \dd t' \left\langle \vct{\omega}(t) \cdot \vct{L} \, \ee^{\ii \Omega (t - t') L_z} \vct{\omega}(t') \cdot \vct{L} \right\rangle \\
&= - \int_{-\infty}^t \dd t' \, \langle \omega_i(t) \omega_j(t') \rangle L_i \ee^{\ii \Omega (t - t') L_z} L_j \, .
\end{align}

For a stationary random field, the two-point function only depends on the difference of times. The turbulent magnetic field along the particle trajectory, $\vct{\omega}(t)$, is a stationary random field, and so we define $C_{ij}(\tau) \equiv \langle \omega_i(t) \omega_j(t') \rangle$ with $\tau \equiv (t - t')$. We assume that $C_{ij}(\tau) \equiv C(\tau) \delta_{ij}$ if $i=x,y$ and zero otherwise, as is appropriate for slab turbulence. Making use of the commutation relations for the angular momentum operators, we can then write for $\tilde{V}(0)$,
\begin{align}
\tilde{V}(0) &= -\int_{0}^{\infty} \mathrm{d}\tau \, C(\tau) \left[\cos{\Omega\tau}(\vct{L})^{2} + \ii \sin{\Omega\tau}{L}_z-\cos{\Omega\tau}({L}_z)^{2}\right] \\
&= - \nu \left[ (\vct{L})^{2} - ({L}_z)^{2} \right] - \ii \Delta \Omega L_z \, .
\end{align}
For a given turbulence model, the decorrelation rate $\nu$ and the shift in the gyrofrequency $\Delta \Omega$ can be computed explicitly. Note that both the correlation function $C(\tau)$ and therefore also the rate $\nu$ depend on the pitch-angle of the particle, that is the angle between the particle momentum and the $\vct{B}_0$-direction.

Ultimately, we obtain the Bourret propagator,
\begin{align}
\vcenter{\hbox{
\includegraphics[scale=1]{figures/single_double_line}
}} &= \ee^{\ii (\Omega + \Delta \Omega) (t-t_0) L_z} \ee^{- \nu \left[ (\vct{L})^{2} - ({L}_z)^{2} \right] (t-t_0)} \, . \label{eqn:Bourret_propagator}
\end{align}
The action of the first factor of this Bourret propagator is the rotation of the phase-space density around the $\vct{B}_0$-direction at the modified gyrofrequency $(\Omega + \Delta \Omega)$. The second factor describes a combination of diffusion in pitch-angle and gyro-phase. This can be seen by inspection of $\left( (\vct{L})^{2} - ({L}_z)^{2} \right)$ in terms of differential operators,
\begin{align}
(\vct{L})^{2} - ({L}_z)^{2} &= - \left( \frac{1}{\sin\theta} \frac{\partial}{\partial \theta} \left( \sin\theta \frac{\partial}{\partial \theta} \right) + \left( \frac{1}{\sin^2\theta} - 1 \right) \frac{\partial^2}{\partial \varphi^2} \right) \\
&= - \frac{\partial}{\partial \mu} \left[ (1 - \mu^2) \frac{\partial}{\partial \mu} \right] - \frac{\mu^2}{1 - \mu^2} \frac{\partial^2}{\partial \varphi^2} \, ,
\end{align}
where we have introduced the pitch-angle cosine $\mu \equiv \cos \theta$. For the case of a gyrotropic distribution, that is $\partial f / \partial \varphi \equiv 0$, eq.~\eqref{eqn:Bourret_propagator} simplifies to
\begin{align}
\vcenter{\hbox{
\includegraphics[scale=1]{figures/single_double_line}
}} &\to \ee^{- \nu \left[ - \frac{\partial}{\partial \mu} \left( (1 - \mu^2) \frac{\partial}{\partial \mu} \right) \right] (t-t_0)} \, ,
\end{align}
which is the formal solution of the (homogeneous) Fokker-Planck equation,
\begin{equation}
\frac{\partial f}{\partial t} = - \nu \frac{\partial}{\partial \mu} \left[ (1 - \mu^2) \frac{\partial f}{\partial \mu} \right] \, .
\end{equation}

\section{Pitch-angle scattering}
\label{sec:pitch-angle-sacttering}

It is instructive to investigate the structure of the (1a) and (1b) contributions to the mixing matrix further.
Since the two contributions are equivalent under exchange of particles it is sufficient to consider only one. To this end the (1a) contribution, Eq.~\eqref{eqn:def_U1a}, can be written as
\begin{equation}
\begin{aligned}
\langle U^A_{\Delta T} U^{B*}_{\Delta T}\rangle^{(1a)} 
&=-\langle U^A_{\Delta T} U^{B*}_{\Delta T}\rangle^{(0)}\int_{0}^{\Delta T}\mathrm{d}T\int_{0}^{T}\mathrm{d}\tau \int\mathrm{d}^3k
\,g(k_\parallel)\frac{\delta(k_\perp)}{k_\perp}e^{ik_\parallel\mu_A\tau}\\
&\quad\times\left[\cos{\Omega\tau}({L}^{A})^{2} + i\sin{\Omega\tau}{L}^A_z - \cos{\Omega\tau}({L}_z^{A})^{2}\right]\\
\end{aligned}
\end{equation}
with the free propagator $\langle U^A_{\Delta T}U^{B*}_{\Delta T}\rangle^{(0)}$. (Here, we have used the notation $U_{\Delta T} \equiv U_{t_0, t_0 + \Delta T}$ in order to stress that the propagator only depends on the time difference.) 
To investigate the action of this operator on a gyrotropic initial state it is sufficient to consider the part that does not involve $L_z$
\begin{equation}
    \langle U^A_{\Delta T} U^{B*}_{\Delta T}\rangle^{(1a)}_{\text{gyro}} =  
    - 
  \int_{0}^{\Delta T}\mathrm{d}T\int_{0}^{T}\mathrm{d}\tau \int\mathrm{d}^3k
  \,g(k_\parallel)\frac{\delta(k_\perp)}{k_\perp}
  e^{ik_\parallel\mu_A\tau}\cos{\Omega\tau}({L}^{A})^{2}.\\
\end{equation}
When taking the boundaries of the $\tau$-integration to infinity which corresponds to the limit taken in quasi linear theory we can introduce
\begin{equation}
\label{eq:sigma_qlt}
\begin{aligned}
\sigma^2_{A}(\Delta T) &= \frac{1}{\Delta T}\int_{0}^{\Delta T}\mathrm{d}T\int_{-\infty}^{\infty}\mathrm{d}\tau \int\mathrm{d}^3k  \,g(k_\parallel)\frac{\delta(k_\perp)}{k_\perp}
e^{ik_\parallel\mu_{A}\tau} \cos{\Omega\tau} \\
&= \int\mathrm{d}^3k  \,g(k_\parallel)\frac{\delta(k_\perp)}{2k_\perp} \left[\delta(k_\parallel\mu_{A} - \Omega) + \delta(k_\parallel\mu_{A} + \Omega)\right]
\end{aligned}
\end{equation}
where the gyoreresonance function from quasilinear theory is recovered.

Resumming this diagram with higher order contributions the two particle Bourret propagator can be defined as
\begin{equation}
B(\Delta T) = \lim_{n\rightarrow\infty}\Big[ \langle U^A_{\frac{\Delta T}{n}} U^{B*}_{\frac{\Delta T}{n}}\rangle^{(0)} + \langle U^A_{\frac{\Delta T}{n}} U^{B*}_{\frac{\Delta T}{n}}\rangle^{(1a)}+ \langle U^A_{\frac{\Delta T}{n}} U^{B*}_{\frac{\Delta T}{n}}\rangle^{(1b)}\Big]^n.
\end{equation}
When this operator acts on an initial narrow pitchangle distribution
\begin{equation}
\begin{aligned}
\langle f_Af^*_B\rangle_{\mathrm{init}}
&=\delta(\mu_A-\mu_{0})\delta(\mu_B-\mu_{0})\\
&=\sum_{\ell_A}Y_{\ell_A}^0(\vartheta_A)Y_{\ell_A}^{0*}(\vartheta_0)\sum_{\ell_B}Y_{\ell_B}^0(\vartheta_B)Y_{\ell_B}^{0*}(\vartheta_0) \, ,
\end{aligned}
\end{equation}
it broadens this narrow pitchangle distribution with the pitchangle diffusion coefficient $\sigma$
\begin{equation}
    \begin{aligned}
         B_{n,\mathrm{gyro}}(\Delta T) \langle f_Af^*_B\rangle_{\mathrm{init}} =& \lim_{n \rightarrow\infty} \sum_{\ell_A}\left[1-\frac{\Delta T}{n} \sigma_A^2\ell_A(\ell_A+1)\right]^n Y_{\ell_A}^0(\vartheta_A)Y_{\ell_A}^{0*}(\vartheta_0) \\
         &\times \sum_{\ell_B}\left[1-\frac{\Delta T}{n} \sigma_B^2\ell_B(\ell_B+1)\right]^n Y_{\ell_B}^0(\vartheta_B)Y_{\ell_B}^{0*}(\vartheta_0) \\
         = & \sum_{\ell_A} e^{-\ell_A(\ell_A+1)\frac{\sigma_A^2}{2}\Delta T} Y_{\ell_A}^0(\vartheta_A)Y_{\ell_A}^{0*}(\vartheta_0)\\
     &\times\sum_{\ell_B} e^{-\ell_B(\ell_B+1)\frac{\sigma_B^2}{2}\Delta T}Y_{\ell_B}^0(\vartheta_B)Y_{\ell_B}^{0*}(\vartheta_0).
    \end{aligned}
\end{equation}
This is essentially a Gaussian in $\vartheta$ with dispersion that grows linearly with time. Hence, the action of the (1a) and (1b) contributions thus leads to the same pitchangle diffusion as in quasilinear theory.

\section{Relations Needed for the (1c) Diagram}
\label{appendix:1crelations}

Sums over prefactors and Wigner 3j symbols can be combined by shifting the index
\begin{equation}
  \begin{aligned}
    &\sum_{m_0} \sigma_+^2\wignerj{\ell_A}{\ell}{\ell_0}{0}{m}{m_0+1}\wignerj{\ell_B}{\ell}{\ell_0}{0}{m}{m_0+1}+\sigma_-^2\wignerj{\ell_A}{\ell}{\ell_0}{0}{m}{m_0-1}\wignerj{\ell_B}{\ell}{\ell_0}{0}{m}{m_0-1}\\
    =&\sum_{m_0=-\ell_0+1}^{\ell_0+1}(\ell_0(\ell_0+1)-m_0(m_0-1))\wignerj{\ell_A}{\ell}{\ell_0}{0}{m}{m_0}\wignerj{\ell_B}{\ell}{\ell_0}{0}{m}{m_0} \\
    +& \sum_{m_0=-\ell_0-1}^{\ell_0-1}(\ell_0(\ell_0+1)-m_0(m_0+1))\wignerj{\ell_A}{\ell}{\ell_0}{0}{m}{m_0}\wignerj{\ell_B}\ell{\ell_0}0m{m_0}\\
    =& \sum_{-\ell_0+1}^{\ell_0-1}(2\ell_0(\ell_0+1)-m_0(m_0-1)-m_0(m_0+1))\wignerj{\ell_A}{\ell}{\ell_0}{0}m{m_0}\wignerj{\ell_B}\ell{\ell_0}0m{m_0}\\
    +& (\ell_0(\ell_0+1)-\ell_0(\ell_0-1))\wignerj{\ell_A}\ell{\ell_0}{0}m{\ell_0}\wignerj{\ell_B}{\ell}{\ell_0}0m{\ell_0}\\
    +& (\ell_0(\ell_0+1)-\ell_0(\ell_0-1))\wignerj{\ell_A}\ell{\ell_0}0m{-\ell_0}\wignerj{\ell_B}\ell{\ell_0}0m{-\ell_0}\\
    =& \sum_{m_0=-\ell_0}^{\ell_0}(2\ell_0(\ell_0+1)-2m_0^2)\wignerj{\ell_A}{\ell}{\ell_0}0m{m_0}\wignerj{\ell_B}\ell{\ell_0}0m{m_0},
  \end{aligned}
\end{equation}
and analogously
\begin{equation}
  \begin{aligned}
    &\sum_{m_0} \sigma_+^2\wignerj{\ell_A}{\ell}{\ell_0}{0}{m}{m_0+1}\wignerj{\ell_B}{\ell}{\ell_0}{0}{m}{m_0+1}-\sigma_-^2\wignerj{\ell_A}{\ell}{\ell_0}{0}{m}{m_0-1}\wignerj{\ell_B}{\ell}{\ell_0}{0}{m}{m_0-1}\\
    =&\sum_{m_0=-\ell_0+1}^{\ell_0+1}(\ell_0(\ell_0+1)-m_0(m_0-1))\wignerj{\ell_A}{\ell}{\ell_0}{0}{m}{m_0}\wignerj{\ell_B}{\ell}{\ell_0}{0}{m}{m_0} \\
    -& \sum_{m_0=-\ell_0-1}^{\ell_0-1}(\ell_0(\ell_0+1)-m_0(m_0+1))\wignerj{\ell_A}{\ell}{\ell_0}{0}{m}{m_0}\wignerj{\ell_B}\ell{\ell_0}0m{m_0}\\
    =& \sum_{-\ell_0+1}^{\ell_0-1}(m_0(m_0+1)-m_0(m_0-1))\wignerj{\ell_A}{\ell}{\ell_0}{0}m{m_0}\wignerj{\ell_B}\ell{\ell_0}0m{m_0}\\
    +& (\ell_0(\ell_0+1)-\ell_0(\ell_0-1))\wignerj{\ell_A}\ell{\ell_0}{0}m{\ell_0}\wignerj{\ell_B}{\ell}{\ell_0}0m{\ell_0}\\
    -& (\ell_0(\ell_0+1)-\ell_0(\ell_0-1))\wignerj{\ell_A}\ell{\ell_0}0m{-\ell_0}\wignerj{\ell_B}\ell{\ell_0}0m{-\ell_0}\\
    =& \sum_{m_0=-\ell_0}^{\ell_0}2m_0\wignerj{\ell_A}{\ell}{\ell_0}0m{m_0}\wignerj{\ell_B}\ell{\ell_0}0m{m_0}.
  \end{aligned}
\end{equation}
\end{document}